\documentclass[smallextended]{svjour3}       

\usepackage[margin=1in]{geometry}

\usepackage[utf8]{inputenc}
\usepackage{array}
\usepackage{booktabs}
\usepackage{multicol}
\usepackage{multirow}
\usepackage{tabularx}
\usepackage{longtable}
\usepackage{threeparttable}
\usepackage{enumitem} 
\usepackage{balance}
\usepackage{hyperref}
\usepackage{centernot}
\usepackage{amsmath,amssymb,amsfonts}
\usepackage{rotating}
\usepackage{xspace}
\usepackage[many]{tcolorbox}
\usepackage{caption}
\usepackage{subcaption}

\usepackage{blindtext}
\usepackage[noadjust]{cite}
\usepackage{threeparttable}
\usepackage{algorithm}
\usepackage{algorithmic}
\usepackage{soul}
\usepackage{xcolor}
\usepackage{soul}

\newcommand{\sectopic}[1]{\vspace{0.2em}\par\noindent{\textit{\bfseries #1}}}

\newcommand{\rev}[1]{\textcolor{black}{#1}}
\newcommand{\romina}[1]{\textcolor{black}{#1}}

\usepackage[backgroundcolor=white,bordercolor=blue,linecolor=blue]{todonotes}
\newcommand{\kashif}{$\mathcal{K}$\textit{ash}\textit{if}\xspace}
\newcommand{\RICE}{\textsc{Rice\_LRT}\xspace} 
\newcommand{\RICEORG}{\textsc{Rice}\xspace}

\begin{document}
%
\title{Classifier or Prompt: A Case Study on Legal Requirements Traceability}

\author{Romina Etezadi\textsuperscript{1}         \and
        Sallam Abualhaija\textsuperscript{2} \and Chetan Arora\textsuperscript{3} \and 
        Lionel Briand\textsuperscript{1,4}
}

\institute{Romina Etezadi \and Sallam Abualhaija \and Chetan Arora \and  Lionel Briand \at
              \email{retez068@uottawa.ca,              sallam.abualhaija@uni.lu, chetan.arora@monash.edu, \\ lbriand@uottawa.ca}           \\
             \\
            {\textsuperscript{1} School of Electrical Engineering \& Computer Science, University of Ottawa, Canada}\\{\textsuperscript{2} SnT Centre for Security, Reliability, and Trust,  University of Luxembourg, Luxembourg}\\ 
            {\textsuperscript{3} Faculty of Information Technology, Monash University, Melbourne, Australia}\\ 
            {\textsuperscript{4} Lero Research Ireland centre for Software Research and University of Limerick, Ireland }\\
 }

\maketitle
\abstract{

New regulations are continually introduced to ensure that software development complies with ethical concerns and prioritizes public safety. 
A prerequisite for demonstrating compliance involves tracing software requirements to legal provisions. \textit{Requirements traceability} is a fundamental task where requirements engineers are supposed to analyze technical requirements against target artifacts, 
often under a limited time budget. Doing this analysis manually for complex systems with hundreds of requirements is infeasible. The legal dimension introduces additional challenges that increase manual effort. 

In this paper, we investigate two automated solutions based on language models, including large ones (LLMs).
The first solution, \kashif, is a classifier that leverages sentence transformers and semantic similarity. The second solution, \RICE, prompts a recent LLM based on \RICEORG, a prompt engineering framework.
Using a publicly available benchmark dataset, we empirically evaluate \kashif and compare it against seven baseline classifiers from the literature (LSI, LDA, GloVe, TraceBERT, RoBERTa, and LLaMa).

\kashif can identify trace links with F2 score of $\approx$63\%, outperforming the best baseline by a substantial margin of 21 percentage points (pp) in F2 score.
On a newly created and more complex requirements document traced to the European general data protection regulation (GDPR), \RICE outperforms \kashif and baseline prompts in the literature by achieving an average recall of 84\% and F2 score of 61\%, improving the F2 score by 34 pp compared to the best baseline prompt.
Our results indicate that requirements traceability in legal contexts cannot be adequately addressed by techniques proposed in the literature that are not specifically designed for legal artifacts. Furthermore, we demonstrate that our engineered prompt outperforms both classifier-based approaches and baseline prompts. 

}


\section{Introduction}\label{sec:introduction}

Technological advancements are significantly transforming software development across diverse domains, such as healthcare~\cite{Caruana:15}. Software applications and automated assistants have become integral to our daily lives~\cite{Zhan:22}. This evolution, driven by recent breakthroughs in artificial intelligence (AI), has led to increasing complexity in software systems~\cite{ahmad2023requirements,Feldt:18}. 
As technology advances, regulations are evolving in parallel to ensure that software systems are developed in accordance with ethical and legal standards. For example, the General Data Protection Regulation (GDPR)~\cite{GDPR} has been in effect since 2018 to address concerns regarding privacy and data protection. Although introduced by the European Union (EU), the GDPR has a global effect, affecting organizations (and software) outside the EU that process personal data of EU residents.

Requirements Engineering (RE) plays a pivotal role in this landscape. 
RE is concerned with specifying and maintaining software requirements that outline the properties and functions of a system-to-be~\cite{Pohl:11}. 
Legal compliance of software systems against applicable provisions can be addressed at different stages of software development. One scenario is to explicitly identify legal requirements early during the requirements elicitation phase, answering the question: ``What legal obligations need to be satisfied by the system for it to be compliant?''.  The elicited legal requirements 
can then be integrated into the software development process while maintaining traceability to the source legal provisions. 
As an alternative scenario, requirements engineers may need to verify compliance of existing software systems with legal provisions during the post-deployment stage, as new regulations have become applicable.
In this case, they must answer the question ``Does the system satisfy the regulation?''. To do so, engineers must analyze the regulation, identify the applicable legal provisions, and trace software requirements to those provisions.  
Both alternatives rely on \textit{requirements traceability analysis}, an essential RE activity concerned with identifying and maintaining trace links between requirements and other artifacts throughout the software development lifecycle~\cite{Meyer:22}. \textit{legal requirements traceability (LRT)} is a special case where requirements are traced to provisions in a regulation and is the focus of this paper. 

\begin{figure*}
  \includegraphics[width=\textwidth]{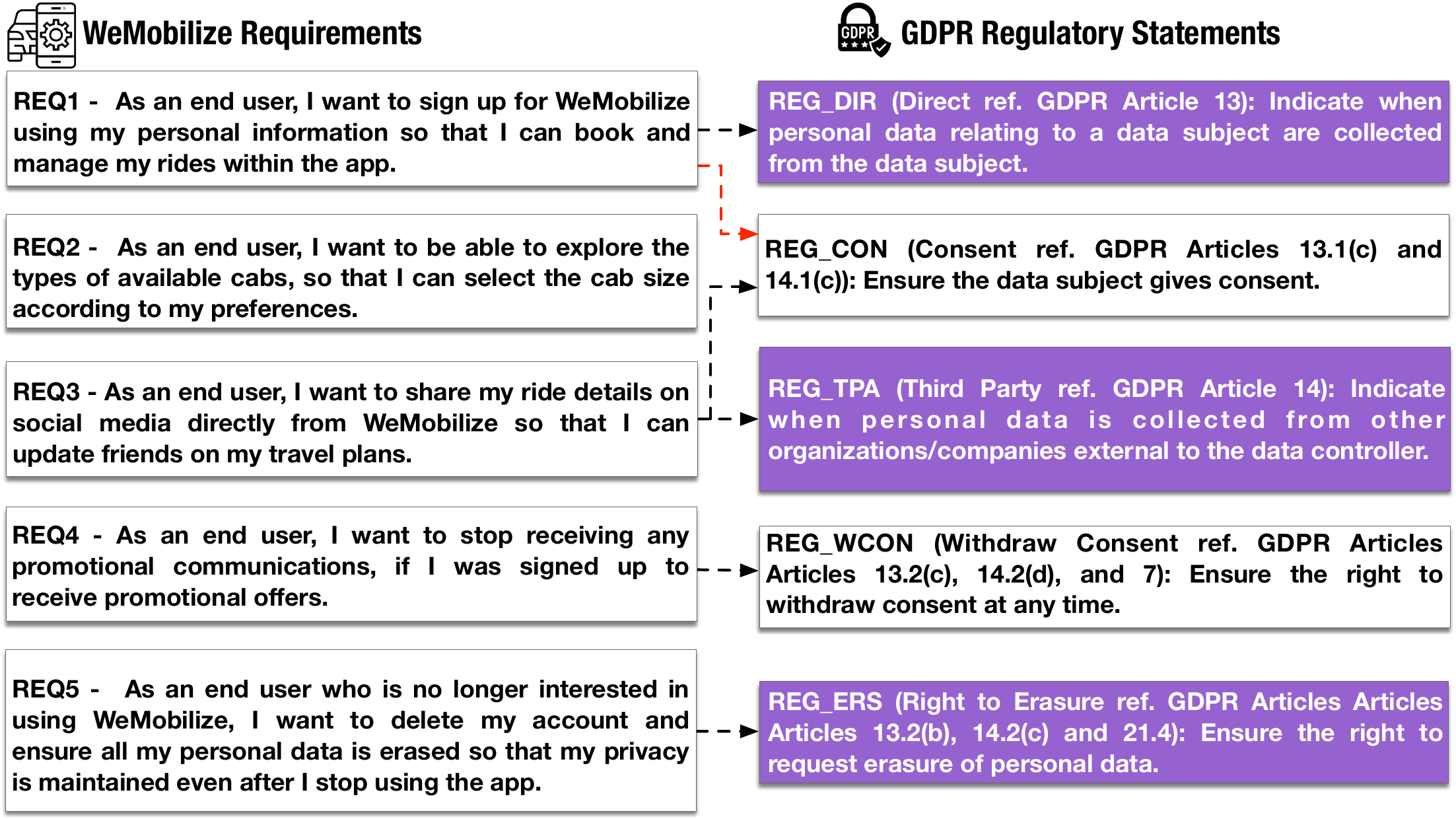}
  \centering
  \caption{Example on tracing WeMobilize app requirements to GDPR statements.}
  \label{fig:example}
 \vspace*{-1em}
\end{figure*}

To illustrate this concept, consider the following example. Consider a fictional mobility app named \textit{WeMobilize}, which helps users book and share taxi rides. Originally a non-EU startup, WeMobilize is expanding into the EU and must therefore comply with the GDPR. 
This example is particularly relevant, as many businesses are globalizing and must adapt to data protection laws across jurisdictions. 
Fig.~\ref{fig:example} illustrates how WeMobilize's requirements (labeled REQ1--REQ5) can be traced to data protection policies under the GDPR~\cite{GDPR}. 
We identify trace links to provisions of the GDPR for REQ1 and REQ3—REQ5, visualized as dashed black lines. REQ2 has no trace link to GDPR in our example, as it does not involve processing users' personal data. 

REQ1 involves collecting users' personal information and must therefore be traced to two provisions: {REG\_DIR} (for the direct collection of personal information) and REG\_CON (for the explicit solicitation of users' consent). Currently, consent is not part of REQ1, which prevents the identification of a trace link with REG\_CON—a missing trace link is visualized as a red dashed line in the figure. Failing to identify this trace link may constitute a breach of the GDPR. Therefore, deploying WeMobilize as-is, without accounting for GDPR provisions, can lead to reputational and financial losses arising from GDPR violations. LRT can help identify potential non-compliance issues at an early stage, but it requires not only legal expertise but also substantial manual effort. Developing automated support is therefore beneficial to assist engineers and analysts in identifying applicable trace links. 

However, achieving this is challenging for LRT due to: 1) \textit{Language Discrepancy:} There is a mismatch in vocabulary, style, structure, and abstraction level between legal regulations and software requirements. Legal text often relies on legal terminology and remains intentionally generic,
while software requirements are often written in technical or domain-specific language.
This discrepancy must be carefully considered when selecting or designing LRT solutions. For instance, REQ4 states, “I want to stop receiving notifications.” Although the term “consent” does not appear explicitly, a human can infer—using common sense and legal understanding—that the ability to stop receiving communications relates to the right to withdraw consent (REG\_WCON). Methods that cannot bridge this conceptual gap may fail to establish such trace links. This highlights the need for models that can reason across domains. 2) \textit{Limited Training Data:} Another major challenge is the scarcity of annotated data for the LRT task. Bridging language discrepancies often requires training modern NLP models on domain-specific examples to effectively capture cross-domain semantics. However, in industrial settings, obtaining large volumes of traced requirement-regulation pairs is difficult due to privacy, confidentiality, or limited documentation. Due to the scarcity of training data, it is crucial to prioritize selecting or designing models that perform robustly with limited examples.

Requirements traceability is a well-explored problem in the RE literature, e.g.,~\cite{wang2018requirements,tufail2017systematic}.
%
%
%
However, the extensive research on requirements traceability is not directly applicable to LRT due to the challenges outlined above.
%
Despite the serious consequences of non-compliance, LRT has received limited attention from the community. 
Cleland-Huang et al.~\cite{cleland:2010,Gibiec:2010} proposed a classifier that predicts trace links by computing the likelihood of requirements being traced to provisions based on indicator terms found in both provisions and requirements. 
Guo et al.~\cite{Guo:17} focused on bridging the terminology gap between provisions and software requirements. 
They examined three methods, including the method by Cleland-Huang et al. mentioned above, as well as two others based on web mining and ontologies. The proposed methods aim to expand the terminology of the provisions by adding additional terms to better identify trace links.

Existing traceability approaches for the LRT task, mentioned above, have several limitations. First, they leverage a statistical model based on TF-IDF (Term Frequency-Inverse Document Frequency), which primarily relies on lexical overlap between documents, where similarity is computed from the presence and frequency of overlapping words. This approach is mostly effective when the source and target texts share similar vocabulary. Still, they struggle to capture deeper semantic relationships, which are required in LRT, as the language used in legal texts often differs significantly from that in technical requirements.
Second, existing techniques have yet to fully leverage the recent advances in natural language processing (NLP), evolving from latent semantic methods such as Latent Semantic Indexing (LSI)~\cite{deerwester1990indexing} and Latent Dirichlet Allocation (LDA)~\cite{blei2003latent}, which uncover hidden semantic structures through concept and topic modeling, toward more contextually rich approaches. We posit that (large) language models (LMs or LLMs) and contextual sentence embedding architectures offer capabilities beyond lexical or shallow syntactic matching, enabling them to capture nuanced legal and technical semantics crucial to LRT. While some recent work has applied language models to traceability tasks~\cite{lin2021traceability}, these efforts have not been tailored to the distinctive demands of aligning with legal requirements—such as interpreting domain-specific terminology, resolving cross-references, or bridging differences in stylistic and structural conventions between legal and technical texts. Architectures like Sentence Transformers can generate semantically rich representations that are better aligned with these needs but remain underutilized in this context. 
%
Third, prompt engineering has not been given sufficient attention; existing work often relies on overly simple prompts~\cite{ronanki2024requirements,hey2025requirements,ge2025cross} or on querying multiple LLMs~\cite{fuchss2025beyond}, which may not always be feasible in real-world deployments due to computational and cost constraints. 
Fourth, the evaluation relies on a single benchmark that may not fully capture the complexity of the legal domain in practice. 
To address these limitations, we propose two novel approaches in this paper, based on recent NLP technologies that utilize the Transformer architecture~\cite{Vaswani:17} and LLMs. Similar to prior work, both approaches aim to predict trace links, and we evaluate their performance in a realistic scenario beyond the commonly used benchmark dataset. 

\sectopic{Contributions.} 
The paper makes the following contributions: 

(1) We devise two automated approaches (Section~\ref{sec:approach}) for predicting trace links between requirements and provisions based on LLMs. 
Our first approach, hereafter referred to as \kashif, standing for \textit{automated trace lin\textbf{K} identific\textbf{A}tion  between legal provi\textbf{S}ions and tec\textbf{H}nical requ\textbf{I}rements using sentence trans\textbf{F}ormers.} 
\kashif leverages sentence transformers (STs), pre-trained language models optimized for understanding longer text sequences such as sentences, and predicts trace links based on semantic similarity. STs are generally more effective than word-level language models~\cite{Reimers:19} for tasks like LRT because they capture the holistic semantic meaning of entire sentences rather than relying on individual word representations. 
While word-level models like BERT, in their default form, produce contextual embeddings for words and require post-processing (e.g., averaging) to represent sentences, they can also produce sentence-level representations through their pooled output (e.g., the [CLS] token) for classification tasks. STs, on the other hand, are a modification of the word-level models that employ siamese and triplet architectures to generate semantically meaningful sentence embeddings~\cite{Reimers:19}, and are explicitly trained on sentence pairs using objectives like contrastive loss. 
This allows them to learn semantic similarity at the sentence level, making them better suited to identifying conceptually equivalent statements—an essential capability in LRT, where the legal and technical texts often have minimal lexical overlap.
Our second approach utilizes \RICE, a recent framework that enables effective prompting of LLMs. We employ \RICE with the GPT-4o model offered by OpenAI\footnote{\url{https://openai.com/index/hello-gpt-4o/}}. Our solutions are described in Section~\ref{sec:approach}.
%

(2) We empirically evaluate our first solution, \kashif, on a benchmark dataset, referred to as \texttt{HIPAA}~\cite{Guo:17}, comprising textual requirements  traced to 10 different provisions. 
We further compare \kashif against a baseline classifier from the literature~\cite{cleland:2010,Guo:17}. Moreover, we compare \kashif with five other baselines employing different technologies, as fully explained in Section \ref{subsec:baseline}.
We re-use, re-implemented, and re-evaluated the baselines as part of this work. 
Our evaluation shows that \kashif yields an average F2 score of $\approx$63\%, leading to a substantial improvement of 34 percentage points (pp) over the best baseline.
While \kashif still performs significantly better than the baselines, such accuracy is rarely practically useful in real-life scenarios where the number of provisions easily exceeds 10 (as is the case in \texttt{HIPAA}). More details on this evaluation can be found in Section~\ref{subsec:rq2}. 

\romina{(3) We created a new dataset based on GDPR regulations. It includes 310 requirements covering diverse domains and requirements types. These requirements are traced to the GDPR, a complex regulation comprising 26 provisions on personal data protection, to which software requirements must comply. }

(4) To further confirm \kashif's performance, we test it on our newly created unseen requirements documents. 
On this dataset, the base ST model (without additional fine-tuning)
yields an average recall of 15\%. In comparison, a pre-trained sentence transformer, with no exposure to the requirements traceability task, yields a nearly zero recall, as elaborated in Section~\ref{subsec:rq3}. 
Driven by this observation, we propose our second solution, the final contribution of this paper, as explained next. 


(5) We devise a prompt strategy based on the \RICEORG framework, capturing recent state-of-the-practice in LLMs for RE. For simplicity, we refer to our prompt strategy hereafter as \RICE. 
Our evaluation (reported in Section~\ref{subsec:rq4}) shows that using \RICE with the GPT-4o LLM leads to an average accuracy of 84\% in predicting trace links in the GDPR dataset, a complex and general regulation. Compared to \kashif, \RICE shows a remarkable gain of 69 pp in accuracy. Moreover, compared to simpler prompts, our engineered prompt outperformed them by 34 pp. 
Using \RICE in practice can significantly reduce the time and effort required to manually identify trace links. With \RICE, the analyst will vet only a small fraction of the provisions, equivalent to $\approx$9.5\%, while identifying 84\% of actual trace links. Further, GPT-4o also provides an informative rationale for each predicted trace link. 



\sectopic{Data Availability. } 
We make our evaluation material available in an online annex~\cite{annex}.

\sectopic{Structure.} Section~\ref{sec:background} provides background. Section~\ref{sec:approach} presents our proposed approaches. Section~\ref{sec:evaluation} reports on our empirical evaluation. Section~\ref{sec:threats} discusses threats to validity. Section~\ref{sec:related} reviews the related work, and finally, Section~\ref{sec:conclusion} concludes the paper.

\section{Background} \label{sec:background}

\sectopic{Language Models (LMs). } Language Modeling in NLP involves computationally determining the probability distribution of word sequences~\cite{Jurafsky:20}. Given a sequence of words, an LM predicts the most likely next word, enabling it to generate text~\cite{Alexandrescu:06}. For example, an LM would predict ``Mat'' as the most likely next word in the input sequence, ``The cat sits on the [WORD]''.  LMs are trained on large corpora of text to accurately estimate these probability distributions. 
State-of-the-art LMs are based on transformer architecture, which leverages self-attention mechanisms to weigh the significance of different parts of an input text relative to a given position~\cite{Vaswani:17}. The attention mechanism determines which words in a sentence are more important in the context and assigns them greater attention. For instance, in the sentence ``Mary, who used to live in Paris, loves wine.'', the attention is on Mary and wine.
Building on transformer architectures, the Sentence Transformers framework (ST)~\cite{Reimers:19} provides a set of pre-trained models that encode longer text sequences, such as sentences or paragraphs, into dense vector representations in a high-dimensional space. They produce contextual embeddings that capture the overall semantic essence of an entire input sequence.

More recently, generative LLMs have emerged as transformer-based language models that are significantly larger and trained on much more data. 
Examples on LLMs include OpenAI's GPT (Generative Pre-trained Transformer)~\cite{Radford:18} and LLaMa~\cite{touvron2023llama,touvron2023llama2}. 
These models can perform new tasks based on textual instructions (prompts)~\cite{NEURIPS2020_1457c0d6}.
%

\sectopic{Machine learning (ML). }
Supervised learning is one of the most prominent paradigms in ML. In this paradigm,  the ML algorithm is provided with labeled training data where each data point consists of an input vector (features) and the corresponding output label (or value). The algorithm learns patterns in the input features to make predictions from this training data. When trained on a sufficiently large dataset, the algorithm refines its predictions to classify the provided labels more accurately. Examples of ML classification algorithms include random forests, decision trees, support vector machines, and feedforward neural networks~\cite{Goodfellow:16}.

\section{Approaches}~\label{sec:approach}
This section defines our notation and then presents our proposed approaches, \kashif and \RICE.

\subsection{Notation}~\label{subsec:def}
Let $\mathcal{R}=\{r_1, r_2, \ldots, r_n\}$ be a set of requirements and $\mathcal{C}=\{c_1, c_2, \ldots, c_m\}$ be a set of provisions derived from applicable regulations. 
Candidate trace links can be created through the Cartesian product between $\mathcal{R}$ and $\mathcal{C}$. 
LRT is then defined as the task of classifying the candidate links into trace links (denoted as $\rightarrow(r_i, c_j)$) or not trace links (denoted as $\centernot\rightarrow(r_i, c_j)$). 



To predict trace links between requirements and provisions, \kashif utilizes Sentence Transformers (ST) and cosine similarity~\cite{Manning:08}.  
%

\subsection{\kashif}
\begin{figure}[t]
\includegraphics[width=0.8\textwidth]{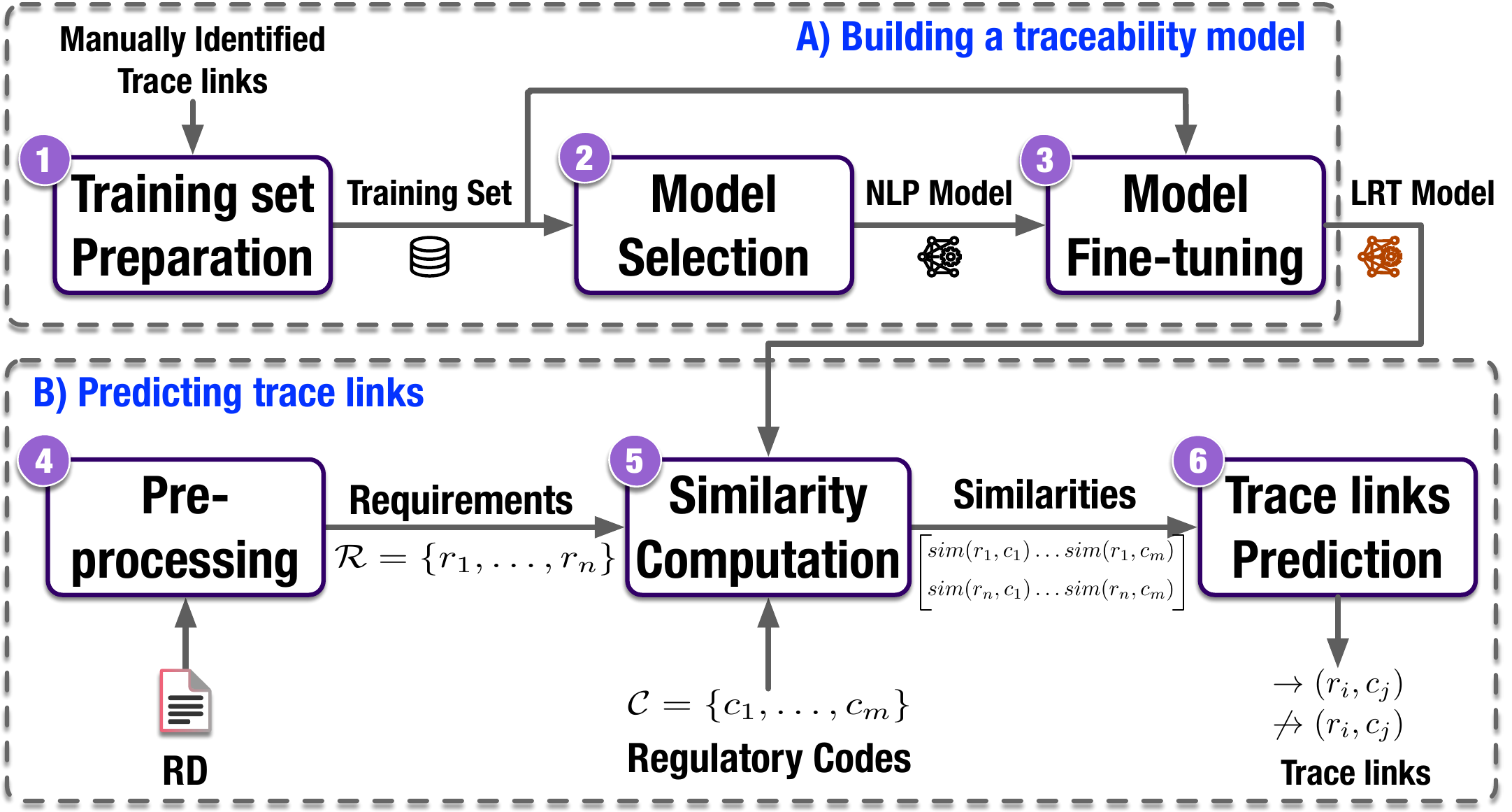}
  \centering
  \caption{Overview of \kashif.}
  \label{fig:approach}
\end{figure}

Fig.~\ref{fig:approach} provides a comprehensive overview of the two phases comprising our approach. 
Phase A covers steps~1-3 and offers a developer's perspective, focusing on building a traceability model for solving LRT. 
Step~1 prepares a training dataset of manually identified trace links. 
Step~2 selects a pre-trained model to customize for addressing LRT. Step~3 involves fine-tuning the LRT model.   
Phase B covers steps~4-6 and provides the perspective of an end user (e.g., a requirements analyst) assuming the availability of an LRT model.  
Step~4 preprocesses the input requirements document (RD). Step 5 applies the LRT model to compute the semantic similarity between each requirement in the RD and each provision. Step~6 predicts trace links. 
We explain these steps in detail next.     

\subsection*{Step~1: Training set preparation } 
Step~1 assumes the availability of a labeled dataset for LRT. We discuss the dataset used in our work in Section~\ref{subsec:datacol}. 
In this step, we transform the training examples into a format suitable for fine-tuning the pre-trained ST models. Each training example is represented as a triple $\langle r_i,c_j,\ell\rangle$, where $\ell=1$ when $r_i$ and $c_j$ have a trace link (positive sample) and $\ell=0$ (negative sample) otherwise. 

\subsection*{Step~2: Model Selection} 
Defining which pre-trained models to start with has become a challenging task due to the regular release of new models\footnote{As of May 15, 2024, there are 124 ST pre-trained models available on HuggingFace.}. 
Ideally, one should fine-tune all available models to select the best-performing one. However, since fine-tuning is resource-intensive, we narrow down the alternatives for experimentation in this step. 
Selecting the best ST model in step~2 is the subject of RQ1,  elaborated in Section~\ref{subsec:rq1}.  

\subsection*{Step~3: Model fine-tuning } \label{sec:fine-tunning}
In step~3, we fine-tune the model selected in step~2. Fine-tuning involves exposing the model to domain-specific knowledge from the provisions and requirements, as well as the particularities of the LRT task. 
During the fine-tuning stage, all requirements in the training documents are considered, regardless of whether they are linked to a provision. We use all possible requirement–provision pairs in the train documents, each labeled with a binary indicator: 1 for positive pairs ($\rightarrow(r_i, c_j)$) and 0 for negative pairs ($\centernot\rightarrow(r_i, c_j)$). Then, the model encodes each text in the pair independently, after which the cosine similarity is computed between their embeddings. To optimize the model, we employ a cosine similarity loss function, which encourages the predicted similarity between a requirement–provision pair to match its true label (1 for linked pairs, 0 for unlinked pairs). This optimization enforces higher similarity scores for positive pairs while penalizing similarity in negative pairs. Negative sampling, grounded in the principles of contrastive learning, enables the model to bring semantically similar textual pairs (requirements and provisions with a trace link) closer in the embedding space while pushing apart dissimilar pairs (requirements and provisions without a trace link). 

The resulting \textit{LRT model} is then passed on to step~5.

\subsection*{Step~4: Preprocessing } 
In  step~4, we preprocess the input requirements using a simple NLP pipeline composed of two modules:  \textit{Tokenization} and \textit{Sentence Splitting}. The goal is to decompose the text into separate sentences. In our work, a requirement $r_i$ corresponds to a sentence generated by the NLP pipeline, which may or may not be grammatically correct. 
Using \kashif to solve LRT for multi-sentence requirements is straightforward. A provision is traced to the requirement if it is traced to any sentence thereof. The intermediary output of this step is a set of $n$ requirements ($\mathcal{R}=\{r_1, r_2, \ldots, r_n\}$) from the input RD. 

\subsection*{Step~5: Similarity Computation }  
Given a set of $m$ provisions $\mathcal{C}$, step~5 computes the semantic similarity scores between each $r_i\in\mathcal{R}$ and each provision $c_j\in\mathcal{C}$. In this work, we apply cosine similarity, which is a widely-used measure for text similarity~\cite{Jurafsky:20}. 
The similarity score is a real value between 0 to 1. A score close to 0 indicates dissimilarity, while a score close to 1 indicates similarity. 
The output of this step is a matrix of dimension $n \times m$, containing the similarity scores between the $n$ requirements in the RD and the $m$ provisions in $\mathcal{C}$. 

\subsection*{Step~6: Trace links Prediction }  
Step~6 predicts a trace link between $r_i$ and $c_j$ using the similarity matrix from step~5. A trace link is predicted when the similarity between $r_i$ and $c_j$ exceeds a certain threshold $\theta$. 
Below, we discuss alternative methods for setting $\theta$. 

\sectopic{(a) Constant Threshold: } 
To predict a trace link, we use a predefined threshold, $\theta = 0.5$. Specifically, a trace link is predicted if the similarity score exceeds $0.5$. 
This threshold is considered a reasonable rule of thumb, as evidenced by its previous application in the literature~\cite{Yao2014,Corley2005}. 
Moreover, this threshold was chosen based on the fine-tuning approach used in \kashif. During training, positive pairs are assigned a similarity score of 1, whereas negative pairs are assigned 0. The model is thus optimized to produce similarity scores closer to 1 for true links and closer to 0 for unrelated pairs. As a result, a threshold of 0.5 serves as the midpoint between these two extremes, making it a natural choice for distinguishing positive from negative links.

\sectopic{(b) Dynamic Threshold: } 
Another practical method to adjust $\theta$ involves curating a set of negative training examples, i.e., requirements that lack trace links. These requirements can be sourced from publicly available datasets or from different projects. However, for more accurate results, it is ideal to use requirements from the same project under analysis. 
Inspired by similarity-based classification proposed in the literature~\cite{Amaral:21}, we select $\theta$ using the following procedure. 
For each provision $c_j \in \mathcal{C}$, we identify a set of negative training examples ($TR_j^-$), i.e., requirements $\{r_1^\prime, \ldots, r_k^\prime\}$ that do not have trace links to $c_j$. We then compute the similarity between $r_i$ and $TR_j^-$ and set $\theta$ to the average cosine similarity between $r_i$ and $TR_j^-$. 
If the similarity between $r_i$ and $c_j$ is higher than the similarity between $r_i$ and $TR_j^-$, then $r_i$ is semantically closer to $c_j$ and should be traced to it.  
Conversely, if the similarity between $r_i$ and $TR_j^-$ is higher, then it should not be traced to $c_j$ as it is semantically closer to the negative examples.   
This procedure sets a different $\theta$ value for each $r_i$ based on randomly selected negative examples. 

\sectopic{(c) Maximum Delta Cutoff: }
In this method, we apply the following procedure. First, for each $r_i$, we sort the similarity values computed across the different provisions $c_j\in\mathcal{C}$. Then, we compute delta values ($\Delta$) corresponding to the differences between each pair of consecutive similarity values and identify the largest $\Delta$ (i.e., the biggest gap in the computed similarities). 
To illustrate, consider the following example. Assume $r_i$ has similarity values of 0.98, 0.1, 0.3, and 0.7 with $c_1$, $c_2$, $c_3$, and $c_4$. We sort these values in descending order as follows: $c_1$: 0.98, $c_4$: 0.7, $c_3$: 0.3, $c_2$: 0.1. Next, we compute the $\Delta$ values: $\Delta(c_1, c_4)$=0.28, $\Delta(c_4, c_3)$=0.4, $\Delta(c_3, c_2)$=0.2. Based on these values, the largest $\Delta$ is 0.4 between $c_3$ and $c_4$. 
Finally, we set $\theta$ to the lower similarity value in the pair that yielded the largest $\Delta$. In the above example, we would set $\theta$ to 0.3 (the similarity value between $r_i$  and $c_3$).  
The largest $\Delta$ represents the most significant drop in similarity, indicating a potential boundary between relevant and irrelevant provision for $r_i$.     
\sectopic{(d) Tuned: } \label{sec:tuned}
In this variation, rather than using a fixed threshold $\theta$, we optimize $\theta$ by performing a search over the training set, evaluating values in the range [0.01, 1) with a step size of 0.01. The optimal threshold with the highest F2 score identified on the training set is then applied to the test set.

The methods described above yield four variants of \kashif, each determined by the value of $\theta$. These variants are referred to as \kashif$_{constant}$, \kashif$_{dynamic}$, \kashif$_\Delta$, and \kashif$_{tuned}$. We compare these variants in  Section~\ref{sec:evaluation}. 

\subsection{\RICE}~\label{subsec:Prompting_LLMs}
\begin{figure}
\includegraphics[width=0.7\textwidth]{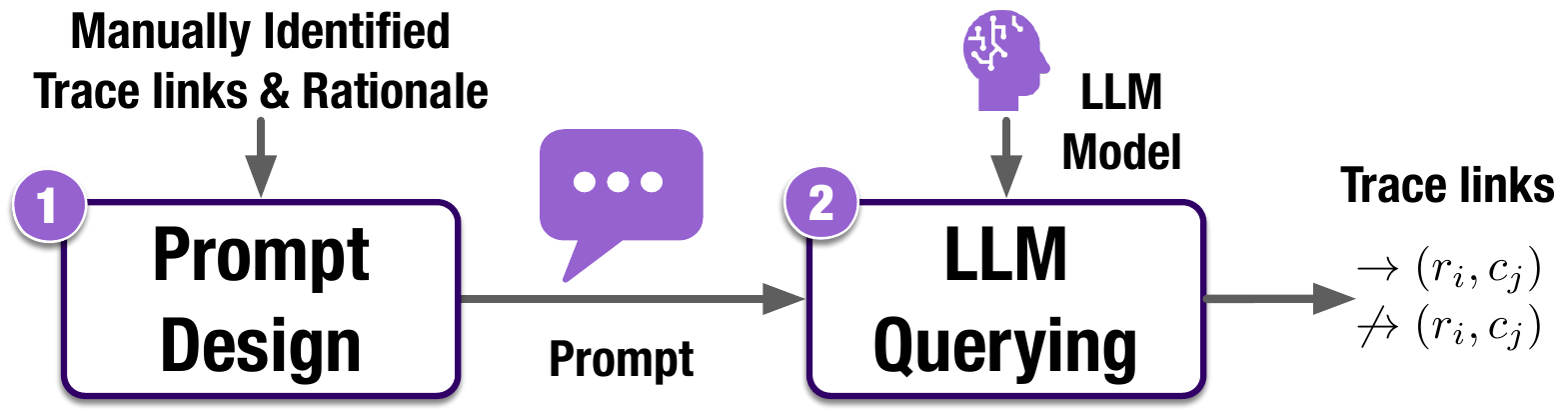}
  \centering
  \caption{Overview of \RICE.}
  \label{fig:rice}
\end{figure}

Our second proposed approach, \RICE, comprises two steps, as illustrated in Fig.~\ref{fig:rice}. The first step is to design a prompt effective for addressing LRT. The second step then applies the prompt to instruct an LLM to predict trace links. We elaborate on these steps next. 

\subsubsection*{Step 1: Prompt Design}
In this step, we designed the prompt in accordance with recent best practices reported in the RE literature~\cite{vogelsang2024using,vogelsang2024specifications}. 
\rev{Fig.~\ref{fig:prompt} presents our final prompt, obtained through iterative refinements. 
The prompt was applied to each requirement in the input RD, with the requirement listed at the end of the prompt. To design the prompt, we followed the \RICEORG (Role, Instruction, Context, Constraints, Examples) framework, with some small adaptations to the LRT task, as we discuss below.
The prompt is structured in the following five elements: }
\begin{itemize}
    \item \textbf{Context:} This element introduces the LRT task. Since the role is implicitly indicated as a requirements analyst building the trace links, this element subsumes the \textit{Role} element in the original 
    \RICEORG framework and simply provides  the \textit{Context}. We omitted the explicit mention of the role to obtain a more general applicability of the prompt. 
    For LRT, multiple analysts with diverse backgrounds are likely involved, for example, a legal analyst in addition to a requirements analyst. 
    \rev{The context corresponds to the text shaded in cyan in Fig.~\ref{fig:prompt}. }
    \romina{\item \textbf{Examples:} This element provides a few examples selected from our ground truth. 
    The examples should cover trace links to different provisions.  Each example comprises a requirement and the set of trace links, along with the rationale for each trace link. We note that the LRT task is complex, as we demonstrate throughout the paper. For this reason, we opted for the few-shot prompting technique. This element matches \textit{Examples} in \RICEORG. Examples corresponds to the text shaded in pink in Fig.~\ref{fig:prompt}.  The few-shot examples are not present in the test data.}
    \item \textbf{Instruction:} This element provides explicit instructions on how to perform the LRT task. This element 
    aims to guide the model through the right reasoning process to generate the desired output. \rev{The \textit{Instruction} element corresponds to the text shaded in olive green in Fig.~\ref{fig:prompt}. }
    Compared with the original \RICEORG framework, this element combines the \textit{Instruction} and \textit{Constraint} elements. This is because the two elements are intertwined in our context. The prompt must therefore account for task-specific considerations, explained below. 
    \begin{itemize}
        \item[$\bullet$] The prompt should encourage the LLM  to equally consider other provisions, since only a subset of the provisions are explicitly explained via the examples and rationales in the \textit{Examples} element. Ideally, the prompt should present an example of each provision. However, this is infeasible, as only the relevant provisions should be traced to the software requirements for a given project. For instance, if the legal basis for collecting personal data is the \textit{contract}, then, unlike explicit consent, only certain \textit{data subject rights} apply under the GDPR and must be appropriately implemented in the software.   
        \item[$\bullet$] \rev{The prompt should account for indirect trace links. As noted above, the LRT is challenging primarily due to the terminology gap between the requirements and the provisions. We therefore encourage the LLM to use its reasoning capabilities to identify indirect links and to generalize beyond the examples provided in the prompt.} 
        \item[$\bullet$] \rev{The prompt should favor recall by predicting at least one trace link for each requirement. 
        As we discuss in Section~\ref{sec:evaluation}, filtering out falsely introduced trace links, as long as they are not too numerous, requires less time and effort by the human analyst than identifying missing trace links.} 
    \end{itemize}
    \item \textbf{Output Indicator:} This element clearly describes the output format, \rev{corresponding to the text shaded in violet in Fig.~\ref{fig:prompt}. }
\end{itemize}
\newcommand{\hlc}[2][yellow]{{%
    \colorlet{foo}{#1}%
    \sethlcolor{foo}\hl{#2}}%
}

\begin{figure}
\begin{tcolorbox}[arc=1mm, width=\columnwidth, 
boxrule=1pt,
colback=gray!15!white,
colframe=black] 
\noindent\textbf{[Context]}\hlc[cyan!50]{\texttt{I am currently working on a task focused on establishing traceability between software requirements and regulatory codes\footnote{\textit{Note that we use regulatory codes to mean provisions, since the former was used in the literature~\cite{cleland:2010}}}. 
This involves analyzing and mapping requirements to relevant GDPR regulations, ensuring that our software development aligns with regulatory compliance. Below are the main regulatory codes that I want you to remember at first: \{The 26 regulatory codes with their descriptions + a 27th code capturing the ``ELSE'' value indicating no trace link.\} }}
\newline 
\textbf{[Examples]} \hlc[pink!50]{\texttt{Here are five sample traceability examples. I've also added my rationale for tracing regulatory codes to the requirements for your reference. \{Five example requirements along with their trace links and the rationale behind selecting these links. Requirement: TEXT. trace links: LIST, rational behind choosing these codes: TEXT.\}}}
\newline 
\textbf{[Instruction]} \hlc[olive!50]{\texttt{Find the trace links for a given requirement and provide the rationale behind your choice extended from the examples I provided. Please consider regulatory codes which I have not used in the examples. Pay attention to the roles (AS\_ROLE) in the requirement, if there are any. Remember, regulations' text focus on personal data, but try to consider all types of data, role, or functionalities in a software system. Pay attention to commonsense and indirect relations between requirement and regulations. Aim to include regulations even if they have a low likelihood of being traced, prioritizing recall over precision. Choose at least one regulation for each requirement.}}
\newline 
\textbf{[Output Indicator]} \hlc[violet!50]{List of alphabetical order of regulatory codes (if any) similar to the examples I provided to you. Newline to explain the rational behind the choice(s).}
\end{tcolorbox}
\caption{Final \RICE prompt for addressing LRT.}
  \label{fig:prompt}
\end{figure}

We note that prompt templates such as \RICEORG serve as illustrative examples or starting points rather than scientifically validated or universally optimal configurations~\cite{vogelsang2024specifications,huang2025prompt}. 
\romina{Therefore, during the refinement process, we randomly selected a separate subset from the few-shot example pool (two out of five examples per document; for a total of eight examples) and used them as inputs for refinement, without including them inside the prompt. In each iteration, a requirement text from this subset was used to obtain the model's response, and the generated rationale was then analyzed to refine the instructions by adjusting constraints, ensuring that the resulting rationales were non-hallucinatory, less ambiguous, and judged as logical by manual evaluation. Note that, after refinement, these examples were returned to the few-shot sets and used as few-shot examples during the test stage.} Moreover, the analysis indicates that presenting the example before the instructions and contextualizing the instructions were more effective than presenting the example at the end.

\subsubsection*{Step 2: LLM Querying}\label{subsec:LLMQuerying}
This step applies the prompt designed in Step~1 to instruct the LLM to predict trace links in textual requirements. 
A prerequisite for using our prompt involves creating a few examples that will demonstrate the LRT task to the LLM. To effectively trigger the reasoning of the LLM, we built five examples by exposing both the labels (i.e., trace links) and the rationale for selecting them. The examples are then integrated into the above LRT prompt template. We applied our prompt by querying the GPT-4o model through the OpenAI API. 

Below, we show an example output for a given requirement from  KeePass, a password management   project\footnote{\url{https://keepass.info/extensions/v1/docs/SoftwareRequirementsSpecification\allowbreak-KeePass-1.10.pdf}} using our prompt \rev{(shown in Fig.~\ref{fig:prompt})}. As mentioned above, {we instruct the LLM through our output indicator to generate the rationale for each regulatory code recommended for a given requirement.} The rationale is meant to help the analysts decide whether to accept or disregard a trace link. 
They can examine the reasoning and logic behind these codes generated by the LLM to understand their potential relevance to the requirement and use their knowledge and interpretation of the requirements documents to decide whether \rev{the proposed trace links by the LLM apply in their context.}

\begin{tcolorbox}[arc=1mm,width=\columnwidth,top=0mm,left=0mm,  right=0mm, bottom=0mm,
                  boxrule=1pt,
               colback=gray!15!white,
               colframe=black, breakable, 
               title=GPT-4o Example Output, breakable]
\texttt{\textbf{REQ:}  \textit{If a key file is required to unlock the database, the database doesn’t open unless the key file is present.}}
\newline
\texttt{\textbf{Trace links:} [ACC, CNF, SEC]}
\newline
\texttt{
\textbf{Rationale:}
\begin{itemize}
    \item ACC\footnote{ACC: Right to access. 
    }: The presence of the key file is required to gain access to the database, ensuring proper authentication.
    \item CNF\footnote{CNF: Ensure confidentiality. 
    }: Protecting the database with a key file ensures that sensitive data remains confidential and accessible only to those with the correct authentication file.
    \item SEC\footnote{SEC: Ensuring security. 
    }: Requiring a key file adds a layer of security, preventing unauthorized access even if the password is compromised.
\end{itemize}
}
\end{tcolorbox}

\section{Evaluation}\label{sec:evaluation}
In this section, we report on our empirical evaluation. 

\subsection{Research Questions (RQs) }
This paper investigates the following RQs: 
\sectopic{RQ1. Which ST model yields the most accurate results for tracing requirements to provisions?}  
As discussed in Section~\ref{sec:approach}, step~2 in \kashif~involves selecting the most accurate pre-trained model for the LRT task. 
Several alternative pre-trained models are publicly available. In RQ1, we examine 38 alternatives reported to work well in the NLP community. The goal of RQ1 is to identify the most accurate ST model for predicting trace links between requirements and provisions. 

\sectopic{RQ2. How accurate is \kashif compared to an existing baseline on a standard dataset from the literature? }
RQ2 aims to assess the value of using ST as an enabling technology for addressing the LRT problem, compared with a baseline from the existing literature that we reimplement in this work. The baseline is a classifier that 
leverages the terminology probability distributions to compute the likelihood that a requirement can be traced to a provision, based on the occurrence of some indicator terms within that provision.
The investigation of RQ2 is conducted using the \texttt{HIPAA} dataset. 

\sectopic{RQ3. How accurately does \kashif perform on a more complex dataset, spanning multiple requirements types and domains?} 
In RQ3, we test \kashif on four different documents, two shall-requirements and two user stories, covering various domains. These documents are derived from the GDPR's privacy requirements. The goal of RQ3 is to evaluate \ kashif's performance on a more realistic dataset that captures the complexity of the legal domain. 

\sectopic{RQ4. How accurate is \RICE in addressing the LRT task compared to \kashif and existing baselines?} 
Given the recent rise in the use of LLMs, a straightforward alternative for automating tasks such as LRT is to prompt pre-trained LLMs, e.g., GPT-4o, for traceability tasks.
RQ4 examines how well our proposed prompt, \RICE, performs compared to the best classifier, \kashif, and existing baselines in the literature.




\subsection{Datasets}\label{subsec:datacol}

\begin{table*}
\footnotesize
\centering
\caption{Statistics of the \texttt{HIPAA} dataset~\cite{cleland:2010}. \rev{Rows list the documents in \texttt{HIPAA}, and columns  provide their description and the distribution of the trace links across provisions in each document. }}\label{tab:hipaa-dataset}
\begin{tabularx}{0.98\textwidth}
{@{} p{0.05\textwidth} @{\hskip 0.5em} p{0.2\textwidth} @{\hskip 0.5em} p{0.05\textwidth} @{\hskip 0.8em} *{11}{>{\centering\arraybackslash}X}@{}}
\toprule
\textbf{ID} & \textbf{Description} & \textbf{All} & \texttt{\textbf{AC}} & \texttt{\textbf{AUD}} & \texttt{\textbf{AL}} & \texttt{\textbf{EAP}} & \texttt{\textbf{PA}} & \texttt{\textbf{SED}} & \texttt{\textbf{TED}} & \texttt{\textbf{TS}} & \texttt{\textbf{IC}} & \texttt{\textbf{UUI}} \\ 
\midrule
H1 & ClearHealth: EMR System. 
& 44 & 1 & 4 & 1 & 0 & 0 & 1 & 1 & 0 & 2 & 1 \\ 
H2 & Physician: Electronic Info. Exchange between Clinicians. 
& 147 & 7 & 2 & 0 & 2 & 0 & 0 & 0 & 1 & 3 & 0 \\ 
H3 & iTrust: Role-based HCT Web app. 
& 184 & 2 & 35 & 1 & 0 & 6 & 0 & 0 & 0 & 0 & 2 \\ 
H4 & Trial Implementations: National Coordinator for Health IT & 100 & 4 & 6 & 0 & 0 & 13 & 0 & 0 & 2 & 4 & 2 \\ 
H5 & PracticeOne: A Suite of HCT Info. Systems. & 34 & 3 & 1 & 0 & 0 & 1 & 0 & 0 & 1 & 1 & 0 \\ 
H6 & PatientOS: HCT Info. System. & 91 & 1 & 2 & 3 & 1 & 0 & 3 & 1 & 1 & 0 & 1 \\ 
H7 & Lauesen: Sample EMR System. & 66 & 11 & 0 & 1 & 0 & 5 & 0 & 0 & 0 & 3 & 1 \\ 
H8 & WorldVistA: Veteran Administrations EMR. & 117 & 6 & 2 & 2 & 0 & 4 & 0 & 0 & 0 & 0 & 1 \\ 
H9 & Care2x: Hospital Info. System. & 44 & 1 & 1 & 1 & 0 & 1 & 1 & 1 & 0 & 0 & 0 \\ 
H10 & CCHIT: Certification Commission for HCT. & 1064 & 17 & 33 & 1 & 1 & 12 & 2 & 2 & 2 & 5 & 3 \\ 
\midrule
& \textbf{Total counts} & \textbf{1891} & \textbf{53} & \textbf{86} & \textbf{10} & \textbf{4} & \textbf{42} & \textbf{7} & \textbf{5} & \textbf{7} & \textbf{18} & \textbf{11} \\ 
\bottomrule
\end{tabularx}
\begin{tablenotes}
     \vspace*{.5em}
 \it    \item[*] 
     \texttt{EMR}: Electronic Medical Record. 
     \texttt{HCT}: Healthcare Technology. 
\end{tablenotes}
\end{table*}

We develop our approach and base our initial evaluation on the \texttt{HIPAA} dataset, a publicly available dataset, created and released in 2010~\cite{cleland:2010} and reused in 2017~\cite{Guo:17}. 
The dataset was manually created by identifying trace links of requirements against the regulatory statements elicited from the USA government's Health Insurance Privacy and Portability Act (\texttt{HIPAA}) regulation. The provisions are the following:  access control (\texttt{AC}), audit controls (\texttt{AUD}), person or entity authentication (\texttt{PA}), transmission security (\texttt{TS}), unique user identification (\texttt{UUI}), emergency access procedure (\texttt{EAP}), automatic logoff (\texttt{AL}), encryption and decryption (\texttt{SED}), encryption (\texttt{TED}), and integrity controls (\texttt{IC}). 
\texttt{HIPAA} comprises 10 requirements documents, all of which are shall requirements, from the healthcare domain. In total, the dataset contains 1,891 requirements, of which 243 have trace links. \rev{Table~\ref{tab:hipaa-dataset} summarizes the different documents (rows) in \texttt{HIPAA}, their description, and the distribution of the trace links across provisions (columns).}

\begin{table*}
\footnotesize
\centering
\caption{Test documents used in RQ3 and RQ4}
\label{tab:test-documents}
\begin{tabularx}{0.98\textwidth}
{@{} p{0.06\textwidth} 
*{1}{>{\arraybackslash}X}@{}
}
\toprule 
ID &  Description (\textbf{S}), Domain (\textbf{D}), Number of requirements (\textbf{N}), Number of trace links (\textbf{T}), Type (\textbf{Y}): (1) ``Shall'' Requirements or (2) User Stories   \\ 
\midrule
RD1 & \textbf{S:} Keepass is about password management, 
 \textbf{D:} cybersecurity, \textbf{N:} 78,  \textbf{T:} 64, \textbf{Y:} 1\\ 
RD2 & \textbf{S:} WASP is about Functionalities and services provided by the WASP platform, \textbf{D:} digital services, \textbf{N:} 69, \textbf{T:} 73, \textbf{Y:} 1\\ 
RD3 & \textbf{S:} Datahub is about information on requirements for data publishers, \textbf{D:}  digital library systems, \textbf{N:} 66, \textbf{T:} 51, \textbf{Y:} 2\\ 
RD4 & \textbf{S:} Scrumalliance is about member interactions and data management on professional networking,  \textbf{D:} professional development and certification systems, \textbf{N:} 97, \textbf{T:} 93, \textbf{Y:} 2\\ 
\bottomrule
\end{tabularx}
\end{table*}



\begin{table*}
\footnotesize
\centering
\caption{Statistics of the Test documents. Columns list the documents and rows provide their the distribution of the trace links across provisions in each document. }\label{tab:traces-dataset}
\begin{tabularx}{\textwidth}
{@{} p{0.2\textwidth} p{0.35\textwidth} @{\hskip 0.4em} *{5}{>{\centering\arraybackslash}X} @{}}
\toprule
\textbf{Provision ID} & \textbf{Description Title} & \textbf{RD1} & \textbf{RD2} & \textbf{RD3} & \textbf{RD4} &\textbf{Total} \\ 
\midrule
\texttt{ACC} & Right to access & 8 & 22 & 27 & 57 & 114 \\ 
\texttt{REC} & Right to rectification & 5 & 2 & 3 & 7 & 17 \\ 
\texttt{RES} & Right to restriction & 0 & 0 & 0 & 0 & 0\\ 
\texttt{CMP} & Right to complaint & 0 & 0 & 0 & 0 & 0\\ 
\texttt{ERS} & Right to erasure & 6 & 2 & 4 & 3 & 15\\ 
\texttt{OBJ} & Right to object & 0 & 0 & 0 & 0 & 0\\ 
\texttt{PRT} & Right to portability & 1 & 0 & 0 & 0 & 1\\ 
\texttt{WCON} & Right to withdraw consent & 0 & 0 & 0 & 0 & 0\\ 
\texttt{CON} & Consent & 0 & 12 & 6 & 10 & 28\\ 
\texttt{CAT} & Personal data category & 0 & 16 & 0 & 0 & 16\\ 
\texttt{SCAT} & Personal data special category & 0 & 0 & 0 & 0 & 0\\ 
\texttt{ORG} & Personal data origin & 0 & 1 & 0 & 0 & 1\\ 
\texttt{DIR} & Direct & 0 & 0 & 0 & 0 & 0\\ 
\texttt{PUB} & Publicly & 0 & 0 & 0 & 0 & 0\\ 
\texttt{TPA} & Third party & 0 & 0 & 0 & 0 & 0\\ 
\texttt{COK} & Cookie & 0 & 0 & 0 & 0 & 0\\ 
\texttt{TEC} & Technical measures & 9 & 0 & 0 & 0 & 9\\ 
\texttt{SEC} & Ensuring security & 25 & 2 & 1 & 1 & 29\\ 
\texttt{SAS} & Security assessment & 1 & 0 & 0 & 0 & 1\\ 
\texttt{TRN} & Personal data transfer & 2 & 0 & 4 & 1 & 7\\ 
\texttt{CHL} & Children & 0 & 0 & 0 & 0 & 0\\ 
\texttt{TIM} & Personal data time stored & 0 & 2 & 0 & 5 & 7\\ 
\texttt{DUR} & Processing duration & 4 & 0 & 0 & 5 & 9\\ 
\texttt{CNF} & Ensure confidentiality & 3 & 5 & 0 & 4 & 12\\ 
\texttt{BRC} & Inform breach to data subject & 0 & 4 & 0 & 0 & 4\\ 
\texttt{NTF} & Data breach notification content & 0 & 5 & 0 & 0 & 5\\ 
\midrule
\textbf{Total Links} && 64 & 73 & 51 & 93 & 281\\ 
\bottomrule
\end{tabularx}
\end{table*}

To address RQ3 and RQ4, we create and curate four documents covering different requirements types and domains. These documents provide a snapshot of a practical scenario that exemplifies the potential complexity of LRT. Moreover, we limit testing of the prompts and LLMs to unseen datasets to minimize the risk of data leakage.
For each document, we manually identify trace links between software requirements and a list of 26 provisions derived from GDPR \rev{and pertinent to software}. Building on existing work~\cite{Amaral:21,Amaral:23a}, the codes were comprehensively created, in collaboration with a legal expert (non-author), to represent the privacy requirements in GDPR pertinent to software engineering. Table~\ref{tab:test-documents} and~\ref{tab:traces-dataset} describe our test documents. 
Two co-authors of this paper, with more than 10 years of expertise in requirements engineering, manually analyzed the four documents and identified the trace links for all requirements. The test documents originate from four independent sources, each from a different domain, as shown in Table~\ref{tab:test-documents}.

It is worth mentioning that the test documents, unlike HIPAA dataset, are newly created resources that have not yet been publicly released. Although the underlying content may not rely entirely on new concepts, the traceability links between the software projects and the GDPR are new and have not been shared before. Therefore, we can confidently claim that there is no data leakage when addressing this research question using GPT4.
%


\subsection{Baselines}\label{subsec:baseline}
To evaluate our proposed solutions, we compare them against carefully selected baselines. We present the baselines for both \kashif and \RICE, including their classifier- and prompt-based variants.

\subsubsection{Classifier Baselines}
For \kashif, we adopt the following approaches from the literature:
(1) well-known information retrieval techniques such as LDA and LSI, (2) a statistical classification approach designed specifically for identifying traceability links in legal documents (we name it \texttt{B} throughout the paper), (3) a more recent static word embedding method, GloVe~\cite{pennington2014glove}, (4) a modern BERT-based model, TraceBERT~\cite{lin2021traceability}, originally developed for detecting trace links between code and issue reports, (5) RoBERTa~\cite{liu2019roberta}, an improved variant of BERT for enhanced language understanding, and (6) a recent technique built upon LLaMa, proposed by Ge et al.~\cite{ge2025cross}, designed to support high-to-low level requirements traceability. This selection ensures a comprehensive comparison across traditional, statistical, static embedding-based, transformer-based, and LLM techniques.

\sectopic{LSI and LDA.}
Vector Space Model (VSM), LSI, and LDA have been widely adopted in the field of requirements engineering, particularly for addressing the traceability problem~\cite{guo2025natural}, which aims to identify and recover links between related software artifacts such as requirements, design documents, and code. VSM represents each document (e.g., a requirement or a provision) as a vector in a multi-dimensional space, where each dimension corresponds to a term from the overall vocabulary. Similarity between two artifacts is typically computed using the cosine of the angle between their corresponding vectors, capturing the degree of term overlap across all artifacts. LSI extends the Vector Space Model by applying Singular Value Decomposition (SVD)~\cite{baeza1999modern} to the term-document matrix, thereby generating a lower-dimensional latent semantic space. The dimensionality of this subspace is controlled by a manually tuned parameter, commonly referred to as k. By capturing the most significant underlying patterns in term usage, LSI implicitly accounts for frequently co-occurring terms, thereby mitigating the vocabulary mismatch problem commonly encountered in information retrieval-based traceability recovery. LDA shares LSI's goal of uncovering the latent semantic meaning in text, but instead of using SVD, it employs a Bayesian model to identify the underlying structure in word usage across documents.

\sectopic{B.}
\rev{We re-implement as part of this work the baseline \texttt{B} from the literature~\cite{cleland:2010,Guo:17}. }
\texttt{B} 
is a probabilistic approach based on the occurrences of words in requirements texts and on how likely these words are to be associated with specific provisions. Specifically, \texttt{B} predicts whether a requirement is traced to a provision by identifying keywords (also known as \textit{indicator terms}) that are present in the requirement. 
Given an input requirement for which the trace link should be predicted, \texttt{B} requires a training set from which likelihood estimates of indicator terms are computed for the input requirement, representing how likely it is to be relevant to a specific regulation.
The training set comprises provisions and software requirements, along with the trace links between them. During training, indicator terms are identified and weighted for each provision by parsing the textual requirements traced to these statements.
The weights are computed based on factors such as term frequency in related requirements, the fraction of regulation-related requirements containing the term, and the fraction of projects (specific to the \texttt{HIPAA} dataset) involving regulation-related requirements that also contain the term. 
\rev{Given the absence of publicly released implementation} for the baseline, we present in this paper a replicated version of \texttt{B} which follows the same procedure described above.  

\sectopic{GloVe.}
We implement a technique based on static word embeddings using GloVe, which generates dense vector representations of words by leveraging global co-occurrence statistics from a large corpus, thereby capturing both semantic and syntactic relationships between words. In our work, we use GloVe as both a trainable embedding layer and the input layer in a simple neural network architecture (with a single-layer architecture to predict the score). In this context, the inputs comprise a requirement and a candidate provision, and the output is a probability score computed via a sigmoid activation function. The model is subsequently fine-tuned on the training dataset, allowing it to adapt the pre-trained word embeddings to the specific characteristics of the traceability task.

\sectopic{TraceBERT.}
we utilize the publicly available implementation of TraceBERT, a transformer-based model that, like our approach, is built on a transformer architecture. In this paper, the authors propose three distinct BERT-based architectures (TWIN, Single, and Siamese) for fine-tuning the pre-trained BERT model for the traceability link-prediction task. Each architecture represents a distinct approach to structuring the input and optimizing the model for semantic similarity between software artifacts. Based on reported evaluation results, the Single-TraceBERT architecture outperforms the other variations, and we adopt it as one of the baseline models in this work.
Note that, unlike the original TraceBERT paper, which includes intermediate training on CodeSearchNet~\cite{husain2019codesearchnet}, we skipped this step as it is not relevant to legal domain traceability as CodeSearchNet contains programming language artifacts which is beneficial for source code and issue traceability, but differs fundamentally—in both terminology and syntactic structure—from establishing traceability between two NL artifacts in our case. Therefore, we directly fine-tuned the authors’ architecture on our dataset.

\sectopic{RoBERTa.}
We adopt the recent BERT-based architecture, RoBERTa, an optimized version of BERT that addresses undertraining issues observed in the original model~\cite{liu2019roberta}. 
In this paper, we use the large version of RoBERTa because of its greater capacity to capture complex language patterns and generate richer contextual embeddings.
The fine-tuning is formulated as a binary classification task, in which RoBERTa is trained on two text inputs (e.g., a requirement and a provision) that are jointly encoded by RoBERTa’s tokenizer, which inserts special tokens to separate them. The model processes the combined sequence, and the hidden state of the [CLS] token is used as the aggregated representation of the text pair. This [CLS] embedding is then passed through a classification head consisting of a linear layer followed by a softmax layer to output the probability distribution over the two classes, 0 (negative pairs, $\centernot\rightarrow(r_i, c_j)$) and 1 (positive pairs, $\rightarrow(r_i, c_j)$). The details of the implementation are discussed in Section~\ref{sec:implementation}.

\sectopic{LLaMa.}
Finally, we utilize the publicly available implementation of a recent approach by Ge et al.~\cite{ge2025cross}, which primarily focused on establishing traceability between low- and high-level requirements using LLaMa2~\cite{touvron2023llama}. In this work, they use data augmentation to fine-tune LLaMa2 at different scales (1.1B, 7B, and 13B) and demonstrate that LLaMa2-7B achieves the best performance. They demonstrate that the fine-tuned LLaMa outperforms both IR- and ML-based approaches across six datasets using a 90/10 train–test split. In our study, we evaluate how well this approach performs on the requirement–provision task under a more complex evaluation setting.

\subsubsection{Prompt Baselines}
For \RICE, we incorporate four baselines, \textit{P1}, \textit{P2}, \textit{P3\_1}, and \textit{P3\_2}, from the literature~\cite{hey2025requirements,ronanki2024requirements}, which focus on identifying links between high-level and low-level requirements. This comparison enables us to assess the effectiveness of our context-rich prompt, \RICE, relative to recent traceability prompt-based techniques reported in the literature.
\sectopic{\textit{P1}.}
As part of this work, we reimplemented an approach based on Retrieval-Augmented Generation (RAG) initially proposed for inter-requirement traceability~\cite{hey2025requirements}. This RAG-based approach consists of two main steps: retrieval and prompting. In the retrieval step, semantic similarity is used to select the top-k target artifacts that are most similar to a given source artifact. For the prompting step, the authors employed two types of prompts: KISS, a simple zero-shot prompt, and CoT, a chain-of-thought prompt.
Based on their results, CoT prompting demonstrated superior performance. Therefore, in this work, we focus on CoT and introduce a minor prompt modification to better suit our LRT task. In our context, the source artifacts are the requirements, and the target artifacts are the provisions, respectively. The prompt used is:

\begin{tcolorbox}[arc=1mm,width=\columnwidth,top=0mm,left=0mm,  right=0mm, bottom=0mm,
                  boxrule=1pt,
               colback=gray!15!white,
               colframe=black, breakable]
\texttt{Below are artifacts from a software system requirement and the GDPR. Is there a traceability link between (1) and (2)? Give your reasoning and then answer with 'yes' or 'no' enclosed in \(<\)trace\(>\) \(<\)/trace\(>\).
\newline
(1) Requirement: \textquotesingle\textquotesingle\textquotesingle{\{source\_content\}}\textquotesingle\textquotesingle\textquotesingle \newline
(2) Regulation: \textquotesingle\textquotesingle\textquotesingle{\{target\_content\}}\textquotesingle\textquotesingle\textquotesingle 
}
\end{tcolorbox}

\sectopic{\textit{P2}.} We adopt the prompt template proposed by Ronanki et al.~\cite{ronanki2024requirements}. Their study presents five templates for identifying traceability between requirements, two of which require human involvement, whereas this paper does not. From the remaining three templates, we select the one that achieved the highest F2 score. We additionally introduce minor modifications to tailor the prompt to the specific needs of our LRT task. The prompt used is as follows:

\begin{tcolorbox}[arc=1mm,width=\columnwidth,top=0mm,left=0mm,  right=0mm, bottom=0mm,
                  boxrule=1pt,
               colback=gray!15!white,
               colframe=black, breakable]
\texttt{Act as a requirements engineering domain expert and list the IDs of the GDPR regulations that are dependent on the following requirement:
\newline
Requirement: \textquotesingle\textquotesingle\textquotesingle{\{content\}}\textquotesingle\textquotesingle\textquotesingle\newline
List of Regulations: \textquotesingle\textquotesingle\textquotesingle{\{List\}}\textquotesingle\textquotesingle\textquotesingle
}
\end{tcolorbox}

\sectopic{\textit{P3}.} We adopt two prompt templates used by Ge et al.~\cite{ge2025cross} as part of their investigation into high- and low-level requirement traceability. Their study evaluated four zero-shot prompt variations, ranging from a simple prompt to a more detailed CoT formulation. In order to have a variety of prompts in our baselines, we employ the prompt with the best results in their work, which we call \textit{P3\_1} throughout the paper,  along with the CoT-based prompt, as it includes more detailed instructions compared to \textit{P1} and \textit{P2}, which we call \textit{P3\_2}. We modify the prompt to better fit the LRT task. The prompts we used are as follows:

\begin{tcolorbox}[arc=1mm,width=\columnwidth,top=0mm,left=0mm,  right=0mm, bottom=0mm,
                  boxrule=1pt,
               colback=gray!15!white,
               colframe=black, breakable]
\texttt{\textit{P3-1:} Consider the following scenario where the Requirement text represents requirements of the software system and Regulation represents a GDPR regulation. Let's think step by step: 1. Identify the key elements of the requirement; 2. Match these elements with the regulation; 3. Highlight any gaps; 4. Infer implicit traceability; Finally, answer with Yes or No if there is a traceability relationship, and provide a brief rationale in about 10 words.
\newline
Requirement: \textquotesingle\textquotesingle\textquotesingle{\{content\}}\textquotesingle\textquotesingle\textquotesingle \newline
Regulation: \textquotesingle\textquotesingle\textquotesingle{\{content\}}\textquotesingle\textquotesingle\textquotesingle
}
\end{tcolorbox}

\begin{tcolorbox}[arc=1mm,width=\columnwidth,top=0mm,left=0mm,  right=0mm, bottom=0mm,
                  boxrule=1pt,
               colback=gray!15!white,
               colframe=black, breakable]
\texttt{\textit{P3-2:} Consider the following scenario where the Requirement text represents requirements of the software system and Regulation represents a GDPR regulation. Answer with Yes or No: Does the regulation trace back to the requirement?
\newline
Requirement: \textquotesingle\textquotesingle\textquotesingle{\{content\}}\textquotesingle\textquotesingle\textquotesingle \newline
Regulation: \textquotesingle\textquotesingle\textquotesingle{\{content\}}\textquotesingle\textquotesingle\textquotesingle
}
\end{tcolorbox}

\subsection{Implementation}\label{sec:implementation}
\sectopic{\kashif.}
We implement \kashif in Python 3.8. For text preprocessing, we use the NLTK toolkit (v. 3.8.1). 
We access the ST pre-trained models through the Hugging Face Transformers library (4.44.0). For fine-tuning, we use the Sentence-Transformers library (2.6.1). We use the same library for computing cosine similarity.  
Our experiments were performed on an RTX 6000 GPU with 24 GB of RAM. For fine-tuning, we relied on the ST29 base model. We performed grid search to tune the hyperparameters, considering epochs=[5, 10, 15], batch size=[8, 16, 32], optimizer=[AdamW, RMSprop], learning rate=[2e-5, 2e-3, 2e-2], weight decay=[2e-3, 2e-2, 2e-1], and warmup percentage=[5e-2, 5e-1]. The best-performing hyperparameters were epochs=5, batch size=8, optimizer=AdamW, linear learning rate=2e-5, weight decay=2e-1, and warmup percentage=5e-2. Checkpoints were saved at the end of each epoch, and the final model was selected based on the lowest observed loss.
\sectopic{\textbf{RICE\_LRT}.}
We implement \RICE in Python 3.8. using the OpenAI (1.97.1) API with the following settings: a temperature of 0, a max-token of 2,000, a frequency penalty of 0, a presence penalty of 0, top\_p = 1, and a random seed of 16. 
\sectopic{\texttt{B}.}
We also implement \texttt{B} in Python 3.8. We have used the scikit-learn library (1.7.0) to implement the probabilistic functions. 

\sectopic{LSI and LDA.}
For LSI and LDA, we utilized the scikit-learn library (v1.7.0), setting the number of components $(n_{\text{components}}$) to 50 within a search space of 5, 10, 25, and 50 components.  
For LDA, the Document–Topic Dirichlet Prior ($\alpha$) and Topic–Word Dirichlet Prior ($\beta$) had the values of 2e-2 within a search space of 1e-2, 2e-2, 1e-1, and 2e-1. \romina{The ranges for $\alpha$ and $\beta$ were chosen to encompass commonly used settings in the literature~\cite{uto2017diverse,silva2021topic}, with particular emphasis on values proportional to $1 / n_{\text{components}}$.}
For both models, we employed TF-IDF as the feature weighting scheme using the scikit-learn library (v1.7.0). Our experiments showed that the LSI model achieved optimal performance when stopwords and punctuation were removed. In contrast, the LDA model performed best when stopwords and punctuation were removed, and lemmatization was applied using the NLTK toolkit (v3.8.1). We did not apply minimum cutoff thresholds for vocabulary, as filtering out low-frequency terms could risk omitting information relevant to certain regulatory provisions. For the maximum document frequency cutoff, we experimented with values of 0, 1e-1, and 1. We observed that as the cutoff value increased, the model's performance decreased. Therefore, we decided not to apply a maximum cutoff in our configuration.

\sectopic{GloVe.}
For GloVe, we used it as input to a single-layer neural network with a sigmoid activation function to predict the similarity score implemented in TensorFlow (2.18.0). We performed grid search to tune the hyperparameters, considering epochs=[5, 10, 15], output dimension=[150, 300], max\_length=[50, 100, 150]. The best-performing hyperparameters were using epochs=10, output dimension=300, and max\_length=50.
\sectopic{TraceBERT.}
For TraceBERT, we used the publicly available data-preparation code to tune the model and retained the fine-tuning settings from the shared implementation.

\sectopic{RoBERTa.}
For RoBERTa, the entire fine-tuning setup is implemented using the Hugging Face Transformers library. We performed a grid search to tune the hyperparameters, considering using epochs=[5, 10, 15], batch size=[8, 16, 32], optimizer=[AdamW, RMSprop], learning rate=[1e-5, 2e-5, 1e-3, 2e-3], and weight decay=[0, 1e-5, 1e-3, 1e-2, 1e-1]. The best-performing hyperparameters were epochs=5, batch size=32, optimizer=AdamW, learning rate=2e-5, and weight decay=1e-2.
\sectopic{LLaMa.}
For LLaMa, we used the publicly available implementation presented by Ge et al~\cite{ge2025cross}. We relied on their released code to generate the training set following the procedure outlined in their paper. Since using the same hyperparameters resulted in a single label being produced for all pairs, we performed another round of hyperparameter tuning, considering learning rate=[2e-8, 2e-7, 2e-5, 2e-5, 2e-4, 2e-3, 2e-2] and weight decay=[0, 1e-5, 1e-3, 1e-2, 1e-1], where the best-performing hyperparameters were learning rate=2e-4 and weight decay=1e-5.
\sectopic{\textit{P1}.}
For \textit{P1}, we reimplemented the approach using the authors' settings. The prompt was executed on GPT-4o with a temperature of 0 and a random seed of 133742243. For the RAG step, we used the same embeddings as the authors proposed, specifically the OpenAI embedding text-embedding-3-large.
\sectopic{\textit{P2}.}
For \textit{P2}, we reimplemented the approach using the authors' settings. While the original paper used GPT-3.5-turbo, our comparative experiments revealed that GPT-4o yielded better outcomes. Therefore, the results for \textit{P2} presented in this paper are based on the GPT-4o model. Moreover, the choice of random seed and temperature was based on the best-performing settings reported in their paper.
\sectopic{\textit{P3}.}
For \textit{P3\_1} and \textit{P3\_2}, we reimplemented the approach using the authors' settings on GPT-4o.

\subsection{Pre-trained Model Selection (RQ1)} \label{subsec:rq1}

\sectopic{Methodology. }  We shortlist the ST models for investigation in our work based on the NLP  leaderboard, which reports the 38 most accurate pre-trained models\footnote{\url{https://www.sbert.net/docs/pretrained_models.html}}. These models have been extensively evaluated for their ability to generate sentence embeddings (i.e., capturing the semantics of the entire text) and for their performance in semantic search (i.e., retrieving relevant answers to a given query). Both tasks closely align with our objectives. 
To identify trace links, we apply the pre-trained models in a zero-shot setting as follows. 
We let each model compute the similarity matrix equivalent to the output of step~5 in our approach (see Fig.~\ref{fig:approach}). 
We then predict a trace link if the similarity value exceeds 
a predefined threshold. Since zero-shot does not require training, we simply run the pre-trained model on the entire \texttt{HIPAA} dataset.

\begin{table}
\centering
\caption{AUC of ST models for LRT on \texttt{HIPAA} (\textbf{RQ1}) 
}
\label{tab:rq1}
\begin{tabularx}{\textwidth}{@{} p{0.05\textwidth} @{\hskip 0.5em} p{0.05\textwidth} @{\hskip 3em} p{0.05\textwidth} @{\hskip 20em} *{5}{>{\centering\arraybackslash}X}@{}}
    \toprule
    \multirow{1}{*}{$K$\tnote{1}} & \multirow{1}{*}{Model\tnote{2}} & \multirow{1}{*}{Name\tnote{1}} & \multirow{1}{*}{\texttt{AUC}\tnote{1}} & \multirow{1}{*}{$K^\dag$\tnote{1}} \\
    \midrule
1 &   \texttt{ST1}  & \texttt{all-mpnet-base-v2} & 0.744 & 16 \\ 
2 &   \texttt{ST2}  & \texttt{gtr-t5-xxl} & 0.725 & 21 \\ 
3 &   \texttt{ST3}  &\texttt{gtr-t5-xl} & 0.789 & 6 \\ 
4 &   \texttt{ST4}  &\texttt{sentence-t5-xxl} & 0.720 & 22 \\ 
5 &   \texttt{ST5}  &\texttt{gtr-t5-large} & 0.743 & 17 \\ 
6 &   \texttt{ST6}  &\texttt{all-mpnet-base-v1} & 0.712 & 25 \\ 
7 &   \texttt{ST7}  &\texttt{multi-qa-mpnet-base-dot-v1} & 0.688 & 27 \\ 
8 &   \texttt{ST8}  &\texttt{multi-qa-mpnet-base-cos-v1} & 0.603 & 34 \\ 
9 &   \texttt{ST9}  &\texttt{all-roberta-large-v1} & 0.601 & 35 \\ 
10 &   \texttt{ST10}  &\texttt{sentence-t5-xl} & 0.769 & 10 \\ 
11 &   \texttt{ST11}  &\texttt{all-distilroberta-v1} & 0.719 & 23 \\ 
12 &   \texttt{ST12}  &\texttt{all-MiniLM-L12-v1} & 0.729 & 19 \\ 
13 &   \texttt{ST13}  &\texttt{all-MiniLM-L12-v2} & 0.747 & 15 \\ 
14 &   \texttt{ST14}  &\texttt{multi-qa-distilbert-dot-v1} & 0.563 & 36 \\ 
15 &   \texttt{ST15}  &\texttt{multi-qa-distilbert-cos-v1} & 0.640 & 33 \\ 
16 &   \texttt{ST16}  &\texttt{gtr-t5-base} & 0.770 & 9 \\ 
17 &   \texttt{ST17}  &\texttt{sentence-t5-large} & 0.748 & 14 \\ 
18 &   \texttt{ST18}  &\texttt{all-MiniLM-L6-v2} & 0.761 & 11 \\ 
19 &   \texttt{ST19}  &\texttt{multi-qa-MiniLM-L6-cos-v1} & 0.670 & 29 \\ 
20 &   \texttt{ST20}  &\texttt{all-MiniLM-L6-v1} & 0.749 & 13 \\ 
21 &   \texttt{ST21}  &\texttt{paraphrase-mpnet-base-v2} & 0.850 & 2 \\ 
22 &   \texttt{ST22}  &\texttt{msmarco-bert-base-dot-v5} & 0.644 & 32 \\ 
23 &   \texttt{ST23}  & \texttt{multi-qa-MiniLM-L6-dot-v1} & 0.715 & 24 \\ 
24 &   \texttt{ST24}  & \texttt{sentence-t5-base} & 0.726 & 20 \\ 
25 &   \texttt{ST25}  & \texttt{msmarco-distilbert-base-tas-b} & 0.701 & 26 \\ 
26 &   \texttt{ST26}  & \texttt{msmarco-distilbert-dot-v5} & 0.685 & 28 \\ 
27 &   \texttt{ST27}  & \texttt{paraphrase-distilroberta-base-v2} & 0.801 & 4 \\ 
28 &   \texttt{ST28}  & \texttt{paraphrase-MiniLM-L12-v2} & 0.794 & 5 \\ 
29 &   \texttt{ST29}  & \texttt{paraphrase-multilingual-mpnet-base-v2} & \textbf{0.859} & 1 \\ 
30 &   \texttt{ST30}  & \texttt{paraphrase-TinyBERT-L6-v2} & 0.787 & 7 \\ 
31 &   \texttt{ST31}  & \texttt{paraphrase-MiniLM-L6-v2} & 0.770 & 8 \\ 
32 &   \texttt{ST32}  & \texttt{paraphrase-albert-small-v2} & 0.737 & 18 \\ 
33 &   \texttt{ST33}  & \texttt{paraphrase-multilingual-MiniLM-L12-v2} & 0.811 & 3 \\ 
34 &   \texttt{ST34}  & \texttt{paraphrase-MiniLM-L3-v2} & 0.757 & 12 \\ 
35 &   \texttt{ST35}  & \texttt{distiluse-base-multilingual-cased-v1} & 0.349 & 37 \\ 
36 &   \texttt{ST36}  & \texttt{distiluse-base-multilingual-cased-v2} & 0.341 & 38 \\ 
37 &   \texttt{ST37}  & \texttt{average\_word\_embeddings\_komninos} & 0.647 & 31 \\ 
38 &   \texttt{ST38}  & \texttt{average\_word\_embeddings\_glove.6B.300d} & 0.636 & 30 \\ 
\bottomrule
\end{tabularx}
\begin{tablenotes}
     \item[1] $K$: The average performance ranking of the models, as reported in the NLP community. $K^\dag$: The ranking of the models based on AUC values computed on \texttt{HIPAA} ($K=1$ indicates the highest AUC). 
      \item [2] \texttt{ST1}--\texttt{ST38} correspond to the models reported at this link (sorted by average accuracy in descending order):     \url{https://www.sbert.net/docs/pretrained_models.html}. 
     \end{tablenotes}
 \end{table}

\sectopic{Evaluation Metrics. } To better assess the performance irrespective of the selected threshold, we compute the \textit{Area Under the Curve (AUC)} for the receiver operating characteristic (ROC) across different threshold values,  ranging from $0.1$ to $0.9$. 
The ROC curve captures the trade-off between the true positive rate (TPR) and the false positive rate (FPR). TPR is the proportion of positives correctly identified as such (i.e., the percentage of trace links correctly identified for a given threshold). FPR is the proportion of negatives incorrectly identified as positives (i.e., the percentage of trace links wrongly identified as not trace links). The AUC of the ROC curve (computed as a micro-average across all provisions to avoid the dominance of any single provision) provides a single aggregate performance measure across all possible thresholds and, hence, is a suitable evaluation metric for comparing ST models.  We posit that the model with the highest AUC performs best in identifying trace links in a zero-shot setting, as a higher AUC indicates a better balance between correctly identifying true trace links (high TPR) and minimizing false positives (low FPR). 
\sectopic{Results. }
Table~\ref{tab:rq1} presents the \texttt{AUC} values of the ST pre-trained models on the \texttt{HIPAA} dataset and also reports $K$, indicating the ranking of the models in the NLP community based on their accuracy~\cite{Reimers:19}, as well as $K^\dag$, indicating the ranking based on \texttt{AUC} achieved on \texttt{HIPAA}.

The best-performing model on \texttt{HIPAA} is \texttt{ST29} ($K^\dag=1$), with an AUC value of 0.859. The next best-performing model is \texttt{ST21}, with an AUC of 0.850. The difference between these two AUC values is only marginal. A possible explanation is that  \texttt{ST29} uses  \texttt{ST21} as its base model.  \texttt{ST29}  has been, however, trained on more (multi-lingual) data.   

Additionally, we observe a discrepancy between the models' performance on the \texttt{HIPAA} dataset and that reported by the NLP community.  
The best NLP model, \texttt{ST1}, does not perform well on \texttt{HIPAA}, ranking 16th. 
This observation indicates that high-performing NLP models are not necessarily effective for RE-specific problems. 
We acknowledge that zero-shot performance does not always predict the effectiveness of fine-tuning, particularly across different model architectures. To address this concern, we fine-tuned the top five models based on their zero-shot AUC scores to evaluate their performance more comprehensively. However, our experiments show that \texttt{ST29} continues to outperform the other models after fine-tuning. While resource constraints prevented us from fine-tuning all possible models, this targeted evaluation provides evidence that \texttt{ST29} is the most effective choice among the top-performing candidates.

\begin{tcolorbox}[arc=1mm,width=\columnwidth,
                  top=0mm,left=0mm,  right=0mm, bottom=0mm,
                  boxrule=1pt, colback=violet!15!white,colframe=white]
\textbf{The answer to RQ1} is that \texttt{ST29} is the best-performing pre-trained model for LRT (corresponding to \texttt{paraphrase-multilingual-mpnet-base-v2}). 
\end{tcolorbox}

\subsection{Accuracy on Benchmark Dataset (RQ2) } \label{subsec:rq2}

\sectopic{Methodology.} We compare the four variants of \kashif (explained in Section~\ref{sec:approach}) against LSI, LDA, \texttt{B} from the literature~\cite{cleland:2010,Guo:17}, GloVe, TraceBERT, RoBERTa, and LLaMa. 
We answer RQ2 on the benchmark dataset, \texttt{HIPAA}.  Since \texttt{HIPAA} comprises 10 requirements documents, we apply the leave-one-out (LOO) evaluation method, in which \kashif and baselines are tested on a left-out document each time and trained (or fine-tuned) on the remaining documents to emulate realistic settings.  
However, to ensure a reasonable balance between the training and test sets, we exclude one document (\texttt{CCHIT}, labeled H10 in Table~\ref{tab:hipaa-dataset}) from the LOO process because it contains 1,064 requirements, thereby comprising more than half of the dataset. 


\sectopic{Evaluation Metrics.} We evaluate the four variations of \kashif and the baselines using precision (P), measuring how many trace links identified by the approach are correct; recall (R), measuring how many trace links in our ground truth are correctly identified by the approach; and F2 score, the harmonic mean of precision and recall, with a greater emphasis on recall.
We report the mean and standard deviation across the nine documents. To facilitate comparisons of the models' rankings, we also report Mean Average Precision (MAP) over requirements with at least one trace link to a provision, which measures how well a model ranks relevant items higher than irrelevant ones. We also report the AUC value for the overall performance of the classifiers. 

Moreover, we report the AUC value for each classifier to better evaluate their performance in distinguishing positive and negative pairs across thresholds. Note that the MAP score is computed only for requirements containing at least one true positive, since requirements without any trace links provide no informative contribution to the MAP. In contrast, AUC is applied to all requirements because it enables a more comprehensive assessment of each classifier's ability to distinguish relevant from non-relevant pairs, regardless of the presence or absence of TPs for a particular requirement.

\begin{table*}
\centering
\caption{Accuracy of  \kashif and baselines on \texttt{HIPAA} (\textbf{RQ2}) }
\label{tab:rq2}
\begin{tabularx}{\textwidth}{@{}l*{7}{>{\centering\arraybackslash}X}@{}}

\toprule

& TP & FP & FN  & P & R & F2 & MAP\\
\toprule
\kashif$_\text{constant}$ & 111\scalebox{0.8}{$\pm 12$} & 114\scalebox{0.8}{$\pm 8$} & 54\scalebox{0.8}{$\pm 4$}  & 49.3\scalebox{0.8}{$\pm 13$} & 67.3\scalebox{0.8}{$\pm 18$} & \textbf{62.7}\scalebox{0.8}{$\pm 11$} & \multirow{4}{*}{\textbf{81.4}}\\
\kashif$_\text{dynamic}$ &  122\scalebox{0.8}{$\pm 12$} & 441\scalebox{0.8}{$\pm 46$} & 43\scalebox{0.8}{$\pm 4$}   & 21.7\scalebox{0.8}{$\pm 18$} & 73.9\scalebox{0.8}{$\pm 21$} & 49.0\scalebox{0.8}{$\pm 13$}\\
\kashif$_\Delta$ & 132\scalebox{0.8}{$\pm 12$} & 1531\scalebox{0.8}{$\pm 81$} & 33\scalebox{0.8}{$\pm 2$}  & 7.9\scalebox{0.8}{$\pm 3$} & \textbf{80.0}\scalebox{0.8}{$\pm 12$} & 28.3\scalebox{0.8}{$\pm 4$}\\
\kashif$_\text{tuned}$ & 94\scalebox{0.8}{$\pm 11$} & 44\scalebox{0.8}{$\pm 4$} & 71\scalebox{0.8}{$\pm 3$}  & \textbf{68.1}\scalebox{0.8}{$\pm 19$} & 56.9\scalebox{0.8}{$\pm 17$} & 58.5\scalebox{0.8}{$\pm 12$} \\
\midrule
\texttt{LDA} & 66\scalebox{0.8}{$\pm$11} & 1571\scalebox{0.8}{$\pm$253} & 99\scalebox{0.8}{$\pm$6} & 4.1\scalebox{0.8}{$\pm$14} & 40.0\scalebox{0.8}{$\pm$24} & 14.4\scalebox{0.8}{$\pm$3} & 37.8\\
\texttt{LSI} & 19\scalebox{0.8}{$\pm$2} & 202\scalebox{0.8}{$\pm$31} & 146\scalebox{0.8}{$\pm$11} & 8.5\scalebox{0.8}{$\pm$18} & 11.5\scalebox{0.8}{$\pm$17} & 10.7\scalebox{0.8}{$\pm$13} & 53.5\\
\texttt{B} & 24\scalebox{0.8}{$\pm$2} & 16\scalebox{0.8}{$\pm$1} & 141\scalebox{0.8}{$\pm$10}  & 60.0\scalebox{0.8}{$\pm$20} & 14.5\scalebox{0.8}{$\pm$9} & 17.1\scalebox{0.8}{$\pm$11} & 78.8 \\
\texttt{Glove} & 50\scalebox{0.8}{$\pm$11} & 428\scalebox{0.8}{$\pm$55} & 115\scalebox{0.8}{$\pm$7} & 10.5\scalebox{0.8}{$\pm$8} & 30.3\scalebox{0.8}{$\pm$25} & 22.0\scalebox{0.8}{$\pm$11} & 48.1 \\
\texttt{TraceBERT} & 15\scalebox{0.8}{$\pm$3} & 97\scalebox{0.8}{$\pm$20} & 147\scalebox{0.8}{$\pm$12} & 13.4\scalebox{0.8}{$\pm$6} & 9.3\scalebox{0.8}{$\pm$25} & 9.8\scalebox{0.8}{$\pm$8} & 62.7\\
\texttt{RoBERTa} & 129\scalebox{0.8}{$\pm$7} & 784\scalebox{0.8}{$\pm$39} & 36\scalebox{0.8}{$\pm$6} & 14.1\scalebox{0.8}{$\pm$1} & 78.1\scalebox{0.8}{$\pm$12} & 41.0\scalebox{0.8}{$\pm$1} & 55.9 \\
\texttt{LLaMa2-7b-w/o$*$} & 33\scalebox{0.8}{$\pm$6} & 1366\scalebox{0.8}{$\pm$229} & 129\scalebox{0.8}{$\pm$11}  & 2.4\scalebox{0.8}{$\pm$1} & 20.3\scalebox{0.8}{$\pm$31} & 8.1\scalebox{0.8}{$\pm$5} & 28.3 \\
\texttt{LLaMa2-7b-tuned} & 111\scalebox{0.8}{$\pm$12} & 1713\scalebox{0.8}{$\pm$102} & 51\scalebox{0.8}{$\pm$3}  & 6.1\scalebox{0.8}{$\pm$2} & 67.3\scalebox{0.8}{$\pm$17} & 22.5\scalebox{0.8}{$\pm$7} & 41.4 \\
\bottomrule
\end{tabularx}

    \begin{tablenotes}
         \item[1] $*$: w/o: without fine-tuning, rather applied in zero-shot setting. 
    \end{tablenotes}
 \end{table*}

\sectopic{Results. }
\kashif$_{\text{tuned}}$ and the baseline models (LDA, LSI, \texttt{B}, and GloVe) require a decision threshold, which we tune using the same strategy explained in Section~\ref{sec:tuned}. In contrast, RoBERTa relies on the softmax function inherent to its architecture for final class-label assignment, and tuning the threshold on the output logits for label 1 yielded a lower F2 score. Similarly, the publicly available implementations of TraceBERT and LLaMA output both class probabilities and a final label, and using the label as the final output yielded a higher F2 score. Therefore, these models do not require separate threshold selection. To compute the MAP value for RoBERTa, we use the output logits (scores) corresponding to the [CLS] token for label 1; for TraceBERT and LLaMa, we use the probability outputs.

\begin{figure}
    \centering
\begin{subfigure}{0.8\textwidth}
        \centering
        \includegraphics[width=\textwidth]{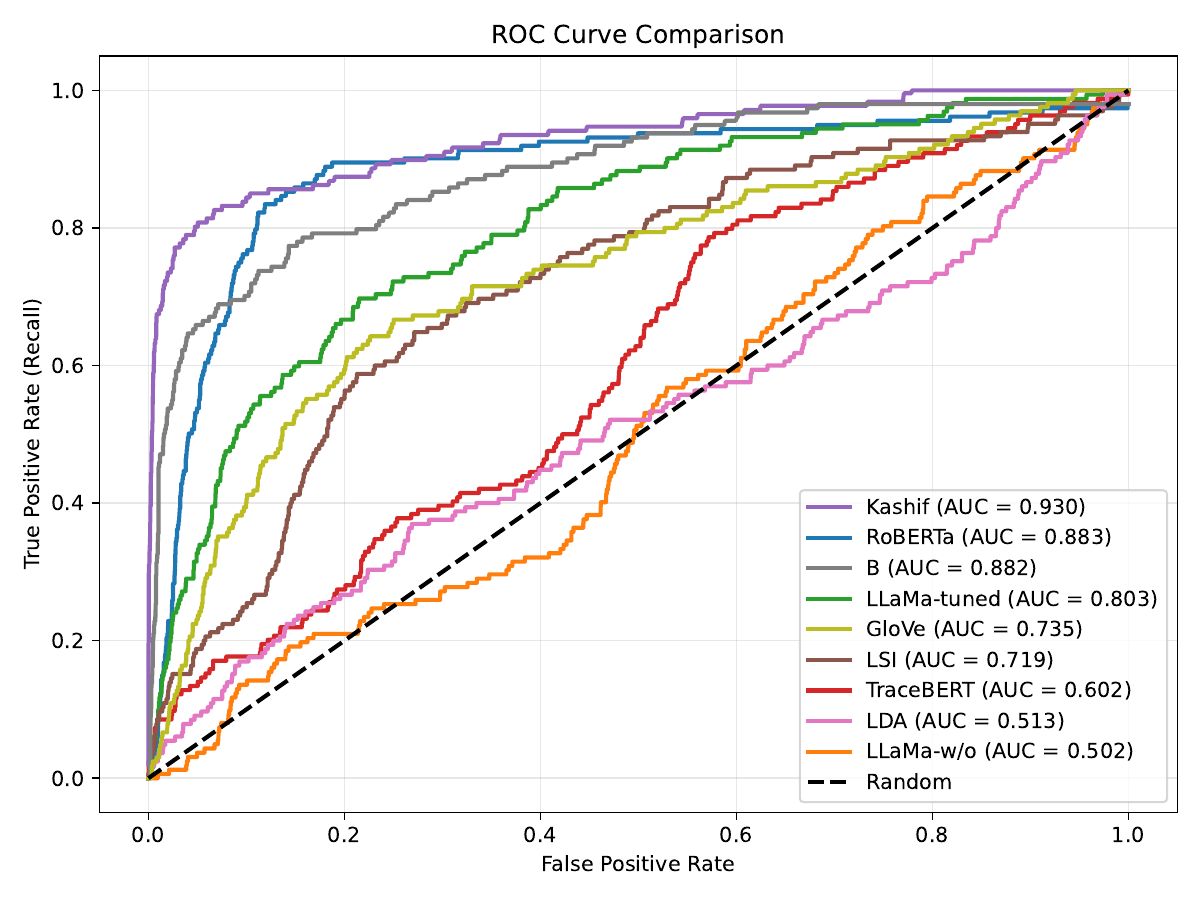}
        \caption{ROC-AUC plot for all classifiers.}
        \label{fig:roc}
    \end{subfigure}
    \begin{subfigure}{0.8\textwidth}
        \centering
        \includegraphics[width=\textwidth]{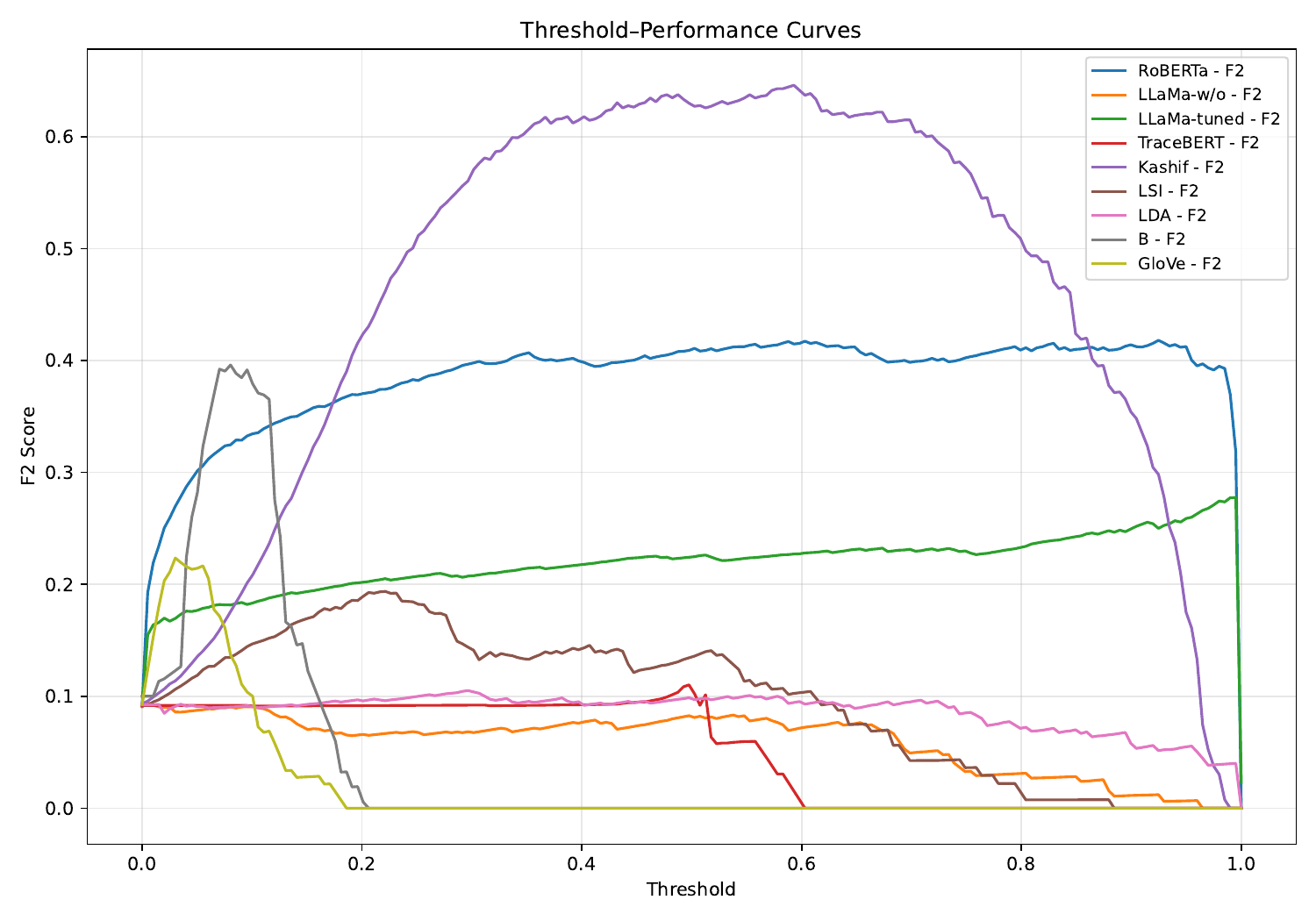}
        \caption{F2 scores over different threshold values. }
        \label{fig:threshold-per}
    \end{subfigure}
    \caption{Evaluating classifier performance across different thresholds.}
    \label{fig:threshold-figures}
\end{figure}

Table~\ref{tab:rq2} lists, for each approach, the total number of TPs, FPs, FNs, and TNs, and further reports the mean and standard deviation of precision, recall, F2, and MAP.
Moreover, Figure~\ref{fig:roc} and Figure~\ref{fig:threshold-per} present the ROC-AUC curve and the average F2 score along 200 evenly spaced threshold values between 0 and 1 for each classifier, respectively.

As visible from Table~\ref{tab:rq2}, \kashif$_{\text{constant}}$ has the highest F2 score compared to the baselines. \kashif also has the highest MAP score (81.4\%), indicating the best ranking performance: in top-k results, it has the most TPs among baselines. Moreover, \kashif has the highest AUC (0.93), indicating greater ability to distinguish between positive and negative pairs. \kashif$_{\text{tuned}}$ outperforms all variants of \kashif in terms of precision, achieving an average of 68.1\%.
Based on our observations of the tuned thresholds, we found that most threshold values exceed 0.5, which explains the higher precision relative to \kashif$_{\text{constant}}$. For the \texttt{H1} to \texttt{H9} projects, the corresponding threshold values are 0.91, 0.5, 0.79, 0.46, 0.82, 0.81, 0.78, 0.73, and 0.58, respectively.

The IR-based methods, LSI and LDA, exhibit notably different behaviors. LDA achieves an F2 score of only 14.4\% and a MAP of 37.8\%, indicating its failure to assign higher ranks to true-positive pairs. Furthermore, its AUC of 0.51 indicates that LDA performs no better than chance and provides little discriminatory power. In contrast, LSI achieves an F2 score of 10.7\%, which is 4 pp lower than that of LDA. However, it outperforms LDA in both MAP (53.5\%) and AUC (0.71). This suggests that LSI has stronger ranking capability. Additionally, Figure \ref{fig:threshold-per} shows that LSI’s maximum achievable F2 score is close to 20\%, whereas LDA’s peak F2 score is around 14\%. Therefore, the threshold-tuning process did not yield the best LSI performance.

\texttt{B}, an ML-based technique tailored to the LRT task, exhibits relatively low retrieval performance compared with the other classifiers. Nevertheless, it achieves the second-highest MAP score (78.8\%), following \kashif, indicating a strong ability to rank positive pairs higher than negative ones. We recall that \texttt{B} is a classifier that primarily uses a probabilistic approach based on word co-occurrence in requirement texts to predict whether each requirement should be traced to a particular provision. This indicates that \texttt{B} has greater discriminative ability between positive and negative pairs than IR-based techniques, GloVe, TraceBERT, and LLaMa. 
Moreover, Figure \ref{fig:threshold-per} shows that \texttt{B}’s maximum achievable F2 score is around 38\%, indicating that threshold tuning was unable to identify an effective decision boundary for model \texttt{B}.

The GloVe classifier achieves an F2 score of 22\%, a MAP of 48.1\%, and an AUC of 0.73. These results indicate that GloVe behaves similarly to LSI. Unlike some other models, threshold tuning for GloVe appears effective, as shown in Figure \ref{fig:threshold-per}. Its maximum attainable F2 score is approximately 22–23\%, and the selected tuned thresholds achieve an F2 score of 22\%.

TraceBERT, however, performs poorly on the LRT task, yielding results comparable to those of LDA. It achieves an F2 score of 8.7\% and an AUC of 0.67, results that are only marginally better than random. However, its relatively high MAP of 62.7\% indicates that it learned the relative similarity between pairs. We observed that TraceBERT produces nearly identical similarity values for most requirement-provision pairs, around 0.5. As a result, its threshold-based classification is substantially hindered. Moreover, in Figure~\ref{fig:threshold-per}, TraceBERT's F2 score starts to change at a threshold value around 0.5. Recall that TraceBERT was originally designed for issue–code traceability and incorporates an intermediate training stage on a large external dataset. Therefore, its inability to achieve high performance in our setting is not unexpected.

RoBERTa achieves an F2 score of 41\%, a MAP of 55.9\%, and an AUC of 0.88, the second-highest AUC among all classifiers. Its high AUC value indicates that RoBERTa effectively separates positive and negative pairs. However, its ranking quality shows greater variability than that of \kashif and \texttt{B}, as reflected in its MAP, which is the fourth-highest overall. RoBERTa exhibits an almost constant F2 score across different threshold values, as shown in Figure~\ref{fig:threshold-per}. The maximum attainable F2 score for RoBERTa is around 41–43\%, suggesting that the softmax-based output layer is already calibrated to produce its best possible performance within this range. 
Among the baselines, RoBERTa achieves the highest F2 score. RoBERTa also achieves the highest recall among all models and \kashif variations, except for \kashif$_{\Delta}$. However, compared to \kashif$_{constant}$, it has a lower recall, which shows weaker effectiveness
in distinguishing relevant links by retrieving many FPs (650 FP links more than \kashif$_{constant}$).

Regarding the use of LLaMa as a classifier, we observe that LLaMa-w/o performs poorly with the lowest F2 score. After fine-tuning, however, the model’s performance improves significantly, demonstrating that it has learned task-specific cues. LLaMa-w/o has the AUC value of 0.50, which shows a random behavior and lacks discrimination on the LRT task. However, LLaMa-tuned achieves an F2 score of 22.5\%, an AUC of 0.80, and a MAP of 41.4\%, indicating that, although the model can distinguish positive from negative pairs (as reflected in its relatively high AUC), its overall ranking and threshold-based performance remain limited. The low MAP value suggests that LLaMa-tuned struggles to consistently assign high similarity scores to all true-positive pairs, capturing only a subset of links while missing many others. Moreover, in Figure~\ref{fig:threshold-per}, the highest attainable F2 score for LLaMa-tuned is above 25\% and lower than 30\%.

According to Figure \ref{fig:threshold-per}, threshold tuning plays a crucial role in achieving optimal F2 scores for classifiers that rely on a threshold for final label assignment. For instance, \texttt{B} reaches an F2 score of only 17.1\% when using a threshold tuned on the training set, even though its maximum attainable F2 score is approximately 38\%. This behavior can also be observed for LSI. In contrast, models such as RoBERTa, which rely on softmax-based probability outputs for label assignment, already operate near their maximum F2 scores, as evidenced in Table \ref{tab:rq2} and Figure \ref{fig:threshold-per}. On the other hand, for \kashif, the optimal F2 scores are primarily concentrated within the 0.4–0.7 threshold range, which is due to its fine-tuning process described in Section~\ref{sec:tuned}. Not only does \kashif achieve the highest F2 score compared to all classifiers, but it also offers the practical advantages of being stable around a 0.5 threshold range and thus reduces the need for threshold tuning. 

Comparing the four variants of \kashif, our results show that \kashif$_{\text{constant}}$ is the best performing variant in terms of F$_2$, achieving an average score of 62.7\%. This score provides a gain of 4.2 pp over \kashif$_{\text{tuned}}$, 13.7 pp over \kashif$_{dynamic}$, and 34.4 pp over \kashif$_{\Delta}$. In terms of recall, however, \kashif$_{\Delta}$ achieves the best value of 80\%, 12.7\% more than \kashif$_{\text{constant}}$. This can be explained by the threshold adjustment method for \kashif$_{\Delta}$. \rev{Recall from Section~\ref{sec:approach} that to determine the threshold above which a trace link is predicted, we look at the largest gap in similarity values between the requirement and the provisions. Once determined, 
\kashif$_{\Delta}$ will always predict at least one trace link for each requirement corresponding to the provision with the highest similarity value that exceeds this gap.  }
Such a method and recall value can indeed be useful when building recommendation systems. However, they come at the cost of increasing FPs (as evidenced by low precision), requiring the human analyst to filter them out. Consequently, we select  \kashif$_{\text{constant}}$ as the best performing model for LRT.

To understand the sources of errors produced by \kashif$_{\text{constant}}$, we analyzed the results per document and provision. The results are listed in Table~\ref{tab:rq2-errors}. Our analysis reveals the following \rev{causes of errors}:

\begin{itemize}
    \item [$\bullet$] \rev{\textbf{Computing low similarity scores for existing trace links. } A majority of FNs (36/54 = 66.7\%) are due to computing low similarity scores between the requirement and the corresponding traced provisions.} 
    These low scores do not exceed the threshold, thus leading to FNs.  
    \item [$\bullet$] \rev{\textbf{Computing high similarity scores when there are no trace links. }} A majority of FPs (96/113 = 84.9\%) are due to falsely predicting a trace link for those requirements that have no trace links in our ground truth. \rev{This case suggests that a binary classifier could help in reducing FPs by predicting whether a requirement should have a trace link or not in the first place. We have conducted several experiments around this hypothesis. Although we observed fewer FPs with a binary classifier, the overall improvement was not statistically significant; therefore, we do not report it in this paper. }
    \item [$\bullet$] \rev{\textbf{Predicting wrong provisions as trace links. }} The remaining FPs and FNs are caused by predicting provisions other than those identified in the ground truth. 
\end{itemize}

\begin{table*}
\small
\centering
\caption{Results of \kashif
($\theta > 0.5$) per document and provision}
\label{tab:rq2-errors}
\begin{threeparttable}[t]
\begin{tabularx}{\textwidth}{@{}*{16}{>{\centering\arraybackslash}X}@{}}
\toprule
& \multicolumn{3}{c}{\texttt{AC}} & \multicolumn{3}{c}{\texttt{AUD}} & \multicolumn{3}{c}{\texttt{AL}} & \multicolumn{3}{c}{\texttt{SED}} & \multicolumn{3}{c}{\texttt{EAP}}  \\
\cmidrule(lr){2-4}\cmidrule(lr){5-7}\cmidrule(lr){8-10}\cmidrule(lr){11-13}\cmidrule(lr){14-16}
 & TP  & FP  & FN  & TP  & FP  & FN  & TP  & FP  & FN  & TP  & FP  & FN  & TP   & FP   & FN  \\
\midrule 
H$_1$  &1 &0 &0 &3 &  12  &1 &1 &0 &0 &1 &1 &0 &0 &0 &0\\
H$_2$  &1 &2 &  6  &1 &0 &1 &0 &0 &0 &0 &0 &0 &1 &0 &1\\
H$_3$  &2 &  9  &0 &35 &4 &0 &1 &0 &0 &0 &  8  &0 &0 &3 &0\\
H$_4$  &1 &2 &3 &6 &0 &0 &0 &0 &0 &0 &0 &0 &0 &0 &0\\
H$_5$    &2 &1 &1 &1 &1 &0 &0 &0 &0 &0 &0 &0 &0 &0 &0\\
H$_6$   &1 &  10  &0 &2 &0 &0 &3 &1 &0 &3 &0 &0 &1 &0 &0\\
H$_7$   &9 &  5  &2 &0 &2 &0 &1 &0 &0 &0 &0 &0 &0 &0 &0\\
H$_8$  &4 &  10  &2 &2 &1 &0 &1 &0 &1 &0 &0 &0 &0 &0 &0\\
H$_9$  &0 &1 &1 &1 &0 &0 &1 &0 &0 &0 &0 &1 &0 &0 &0\\
\midrule 
$\sum$  &21 &40 &15 &51 &20 &2 &8 &1 &1 &4 &9 &1 &2 &3 &1\\
\midrule
& \multicolumn{3}{c}{\texttt{TED}} & \multicolumn{3}{c}{\texttt{IC}} & \multicolumn{3}{c}{\texttt{PA}} & \multicolumn{3}{c}{\texttt{TS}} & \multicolumn{3}{c}{\texttt{UUI}}\\
\cmidrule(lr){2-4}\cmidrule(lr){5-7}\cmidrule(lr){8-10}\cmidrule(lr){11-13}\cmidrule(lr){14-16}
 &  TP   &  FP   &  FN   &  TP   &  FP   &  FN   &  TP   &  FP   &  FN   &  TP  &  FP  &  FN  &  TP  &  FP  &  FN  \\
\midrule 
H$_1$  & 0 & 1 & 1 & 0 & 1 & 2 & 0 & 0 & 0 & 0 & 0 & 0 & 0 & 0 & 1  \\
H$_2$  & 0 & 0 & 0 & 2 & 0 & 1 & 0 & 0 & 0 & 0 &   6  & 1 & 0 & 0 & 0 \\
H$_3$  & 0 & 0 & 0 & 0 & 1 & 0 & 5 & 1 & 1 & 0 & 1 & 0 & 2 & 3 & 0 \\
H$_4$  & 0 & 0 & 0 & 3 & 3 & 1 & 8 & 3 &   5   & 0 & 0 & 2 & 0 & 3 & 2  \\
H$_5$    & 0 & 0 & 0 & 0 & 0 & 1 & 0 & 0 & 1 & 1 & 3 & 0 & 0 & 1 & 0  \\
H$_6$   & 1 & 2 & 0 & 0 & 0 & 0 & 0 & 4 & 0 & 0 & 0 & 1 & 0 & 0 & 1  \\
H$_7$   & 0 & 0 & 0 & 0 & 0 & 3 & 2 & 1 & 3 & 0 & 0 & 0 & 0 & 0 & 1  \\
H$_8$  & 0 & 0 & 0 & 0 & 2 & 0 & 0 & 1 & 4 & 0 & 0 & 0 & 0 & 3 & 1  \\
H$_9$  & 0 & 0 & 1 & 0 & 0 & 0 & 1 & 0 & 1 & 0 & 0 & 0 & 0 & 0 & 0 \\
\midrule
$\sum$  & 1 & 3 & 2 & 5 & 7 & 8 & 16 & 10 & 14 & 1 & 10 & 4 & 2 & 10 & 6 \\
\bottomrule
\end{tabularx}
    \begin{tablenotes}
     \it\item[*] See Table~\ref{tab:hipaa-dataset} for the names of the documents
    \end{tablenotes}
 \end{threeparttable}
 \end{table*}

\begin{tcolorbox}[arc=1mm,width=\columnwidth,
                  top=0mm,left=0mm,  right=0mm, bottom=0mm,
                  boxrule=1pt, colback=violet!15!white,colframe=white]           
\textbf{The answer to RQ2} is that \kashif yields the  best accuracy  on \texttt{HIPAA} when we apply a constant threshold value of 0.5.  Specifically, \kashif achieves an F2 score of $\approx$63\%. Compared with the best existing baseline in the literature, \kashif achieves a gain of approximately 34 points in F2 score. Moreover, \kashif achieves the highest MAP score of 0.81 and AUC value of 0.93 compared to the baseline classifiers.  
\end{tcolorbox}

\subsection{Effectiveness of Classification (RQ3) } \label{subsec:rq3}

%

\sectopic{Methodology.} 
%
\rev{In real-life scenarios, dealing with LRT involves navigating through many provisions, usually significantly more than 10, as in the simple \texttt{HIPAA} case. This inherent complexity is notable with the 26 provisions pertaining to software in the GDPR.}
Using the test documents described in Section~\ref{subsec:datacol}, we evaluate and compare two models, namely \texttt{ST29}---the best pre-trained ST model selected in RQ1  and \kashif$_{\text{constant}}$---the best \kashif variant fine-tuned on \texttt{HIPAA} identified in RQ2. \rev{Note that we opted not to fine-tune \kashif again on the new documents for three reasons. First, the documents are too small to support meaningful training (or fine-tuning). Second, we aim to evaluate existing solutions in a more realistic scenario and apply them to unseen documents. Finally, \kashif is a similarity-based solution that has been exposed to both the LRT task and the regulatory domain (terminology) during the first fine-tuning on \texttt{HIPAA}. Therefore, further fine-tuning is unlikely to add value. }


\sectopic{Evaluation Metrics.} 
To evaluate the models' effectiveness, we report results at the requirements and trace-link levels.

At the requirements level, we report (i) the number of requirements where the recommendations made by the model were exactly the same as our ground truth (\emph{exact match}); (ii) the number of requirements that were a \emph{partial match} to the ground truth, i.e., the requirements where the model recommended the same regulatory codes as in the ground truth along with additional recommendations (FP); 
(iii) the number of \emph{incorrect matches}, i.e., all the other requirements that are not exact or partial matches. 
Following this, we compute the \textit{success rate} as the ratio of requirements for which the approach predicts correct trace links (including both partial and exact matches) to the total number of requirements.

We also aim to evaluate, on average, the model's performance for a single requirement. Thus, we report two additional metrics: (iv) Macro Recall (\emph{MR}), presented in Equation~\ref{eff}, to assess how well the predicted provisions are actually traced to the requirement; (v) \emph{Cost}, presented in Equation~\ref{cost}, to capture the effort required for an analyst to check the predicted results and filter out the FPs.

\begin{equation}\label{eff}
    MR = \frac{1}{N_r} \sum_{i=1}^{N_r} \frac{TP_{r_i}}{TP_{r_i} + FN_{r_i}}
\end{equation}
\begin{equation}\label{cost}
    Cost = \frac{1}{N_r} \sum_{i=1}^{N_r} \frac{TP_{r_i} + FP_{r_i}}{N_{p}}
\end{equation}

$N_r$ denotes the number of requirements in the dataset, $N_{p}$ the total number of provisions in the GDPR, $TP_{r_i}$ and $FP_{r_i}$ the number of true positives and false positives predicted as a trace link for requirement $i$. 
A higher \emph{MR} value indicates a more complete set of links, with the maximum possible value being 1 (100\%). If a requirement has no trace links and the predicted set is also empty, the recall is defined as 1. A lower \emph{Cost} corresponds to more efficient predictions, thereby reducing analyst effort.

At the trace-link level, we report the total number of trace links, true positives (TPs), false positives (FPs), recall (R), precision (P), and F2 score.

\begin{table*}
\footnotesize
\centering
\caption{Accuracy of \texttt{ST29} and \kashif on the test documents (\textbf{RQ3}).} 
\label{tab:rq3}
\begin{threeparttable}[t]
\begin{tabularx}{\textwidth}{@{}l*{13}{>{\centering\arraybackslash}X}@{}}
\toprule
&&&  \multicolumn{10}{c}{\textit{Trace Link Level}} \\ 
\cmidrule(lr){4-13}
&&&\multicolumn{5}{c}{\texttt{ST29}}&\multicolumn{5}{c}{\kashif}\\ 
\cmidrule(lr){4-8} \cmidrule(lr){9-13}
& N & T$^*$ & TP & FP & R & P & F2 & TP & FP & R & P & F2 \\
\midrule
RD1 & 73 & 57 & 0 & 1 & 0.0 & 0.0 & nan & 10 & 95 & 17.5 & 9.5 & 15 \\ 
RD2 & 64  & 65 & 1 & 3 & 0.2 & 25.0 & 24.9 & 11 & 72 & 16.9 & 13.2 & 16 \\
RD3 & 61  & 43 & 0 & 15 & 0.0 & 0.0 & nan & 7 & 69 & 16.3 & 9.2 & 14.1 \\
RD4 & 92 & 86 & 2 & 1 & 0.1 & 66.7 & 12.5 & 8 & 94 & 9.3 & 7.8 & 8.9 \\
\midrule
Average & - & - & - & - & 0.1 & 22.9 & 9.3 & - & - & 15 & 9.9 & 13.5\\
\midrule
\end{tabularx}

\begin{tabularx}{\textwidth}{@{}l*{13}{>{\centering\arraybackslash}X}@{}}
&&&  \multicolumn{10}{c}{\textit{Requirement Level}}\\ 
\cmidrule(lr){4-13}
&&&\multicolumn{5}{c}{\texttt{ST29}}&\multicolumn{5}{c}{\kashif} \\ 
\cmidrule(lr){4-8} \cmidrule(lr){9-13}
& N & T$^*$ & EM & PM & SR & MR & C & EM & PM & SR & MR & C \\
\midrule
RD1 & 73 & 57 & 32 & 1 & 45.2 & 0 & 0.09 & 19 & 16 & 47.9 & 10.2 & 5.1 \\ 
RD2 & 64  & 65 & 30 & 0 & 46.9 & 0.9 & 0.02 & 29 & 5 & 54.7 & 15.3 & 4.6 \\
RD3 & 61  & 43 & 23 & 4 & 44.3 & 0 & 0.1 & 13 & 16 & 47.5 & 18.9 & 5.9 \\
RD4 & 92 & 86 & 20 & 0 & 21.7 & 4.2 & 0.03 & 14 & 10 & 26.1 & 7.4 & 3.0\\
\midrule
Average & - & - & - & - & 39.5 & 1.3 & 0.06 & - & - & 44.1 & 12.9 & 4.7 \\
\bottomrule
\end{tabularx}

\begin{tablenotes}
     \it\item[] T$^*$: Predicted trace links, EM: Exact Match, PM: Partial Match, SR: Success Rate, MR: Macro Recall, C: Cost. 
     \end{tablenotes}
 \end{threeparttable}
 \end{table*}

\sectopic{Results. }
Table~\ref{tab:rq3} shows the results for each approach across the test documents, both at the trace link level and at the requirement level. The table shows the number of requirements in each test document\footnote{Note that we leave out five requirements from each document to enable fair comparison with the \RICE approach presented in RQ4, which relies on few-shot learning.}, the number of predicted trace links (T$^*$), TPs, FPs, R, P, and F2. It also includes the number of requirements with exact matches, the number with partial matches, and the success rate. 
From the table, we observe that \texttt{ST29} performs worse than \kashif in terms of success rate. While \texttt{ST29} achieves a higher number of exact matches, it produces fewer partial matches compared to \kashif.
Note that, for both techniques, exact matches often represent ``no trace link'', i.e., not predicting any trace link for requirements that had no trace links according to our ground truth. This highlights \texttt{ST29}'s failure to produce a non-empty set of trace links, resulting in a lower success rate. 
Therefore, allowing more partial matches to achieve a higher success rate, even at the cost of fewer exact matches, ensures broader coverage of correct links. Indeed, the effort required for a requirements analyst to filter out FPs is relatively low compared to the consequences of a requirement not being traced to a complete set of predicted provisions. Therefore, achieving a higher number of partial matches, despite FPs, can be more beneficial than achieving a higher number of exact matches for supporting requirements-level analysis. 
Our results indicate that the ST pre-trained model (\texttt{ST29}) failed to automatically predict trace links in most cases, suggesting that it neither understood the LRT task nor the application domain. Moreover, \texttt{ST29} has the \emph{Cost} of 0.06\% and a \emph{MR} of 1.3\% per requirement. This means that, on average, for a given requirement, the model retrieves 0.06\% of the provisions while ensuring that 1.3\% of the actual trace links are included among those retrieved. \kashif achieves a \emph{Cost} of 4.7\% while with the \emph{MR} value of 12.4\%. Fine-tuning the models improves performance on the test documents; however, the more complex patterns in the unseen test documents are still not fully captured by \kashif.

To summarize, \kashif consistently outperforms \texttt{ST29} across all documents, with a notable difference in the number of partial matches. Results thus suggest that fine-tuning pre-trained models on a dedicated dataset is necessary for the models to learn the LRT task.  However, while better than the pre-trained model, \kashif shows the following limitations: 1) it does not provide a rationale behind selecting a trace link, except for the fact that semantic similarity exceeds a pre-defined threshold. This is expected to impede its practical use.  2) The average success rate achieved by \kashif is about 44\%, which is not particularly effective. 


\begin{tcolorbox}[arc=1mm,width=\columnwidth,
                  top=0mm,left=0mm,  right=0mm, bottom=0mm,
                  boxrule=1pt, colback=violet!15!white,colframe=white]
\textbf{The answer to RQ3 is} that \kashif outperforms \texttt{ST29}, demonstrating that fine-tuning helps the model learn about the LRT task. However, \ kashif's performance still shows significant room for improvement in unseen domains.
\end{tcolorbox}
\subsection{Effectiveness of Large Language Model (RQ4) } \label{subsec:rq4}

The baselines' performance on the LRT task is extremely poor, highlighting the need for improvement (RQ2). When we adopted a more refined approach using the ST models (\kashif), it outperformed the baselines but did not yield satisfactory results on an unseen dataset (RQ3). This indicates that ST models can partially address issues inherent in baseline approaches.
However, the ST models lack the robustness needed to generalize effectively across unseen data, as discussed in RQ3. Given their promising results across many tasks~\cite{AroraHH24,abs_2310_18648}, RQ4 aims to assess whether prompting LLMs with an engineered prompt offers a meaningful alternative for LRT and to compare its effectiveness against prompts that were not engineered. We posit that LLMs, with their pretraining across diverse domains, may significantly improve trace link recovery.

\sectopic{Metholodology.} As discussed in Section~\ref{subsec:Prompting_LLMs}, we designed a prompt, based on the \RICEORG structure~\cite{vogelsang2024using}. We prompted the GPT-4o model to generate recommendations for trace links between requirements and GDPR provisions. We base our analysis on the four documents discussed in RQ3. We compare the LLM's recommendations, generated with our prompt for each requirement, against our ground truth.
We begin by evaluating \RICE under two temperature values (0 and 0.5) and five different random seed values. We then compare its performance with the baseline prompts using their reported best settings.

\begin{table*}
\footnotesize
\centering
\caption{\RICE performance across different seed and temperature values.}
\label{tab:rq4-seed}
\begin{threeparttable}[t]
\begin{tabularx}{\textwidth}{@{}p{0.12\textwidth} @{\hskip 0.1em} l*{24}{>{\centering\arraybackslash}X}@{}}
\toprule
\multirow{2}{*}{\textbf{seed}} && \multicolumn{8}{c}{\textit{Temperature=0}} & \multicolumn{8}{c}{\textit{Temperature=0.5}} \\ 
\cmidrule(lr){3-10} \cmidrule(lr){11-18}
&& TP & FP & F2 & EM & PM & SR & MR & C & TP & FP & F2 & EM & PM & SR & MR & C \\
\midrule
\multirow{5}{*}{0} 
& RD1 & 46 & 108 & 58.4 & 6 & 55 & 83.6 & 77.5 & 7.9 & 43 & 115 & 55.7 & 5 & 54 & 80.8 & 77.0 & 8.3\\ 
& RD2 &  52 & 134 & 58.3 & 2 & 53 & 85.9 & 83.8 & 11.1 & 49 & 126 & 56.3 & 4 & 46 & 78.1 & 72.5 & 10.5\\
& RD3 & 38 & 112 & 59.0 & 2 & 54 & 91.8 & 88.6 & 9.4 & 35 & 111 & 55.0 & 3 & 50 & 86.9 & 80.0 & 9.2\\
& RD4 & 76 & 150 & 66.7 & 6 & 79 & 92.4 & 92.1 & 9.4 & 74 & 149 & 65.3 & 8 & 75 & 90.2 & 89.3 & 9.3\\
\cmidrule{2-18}
&Average & - & - & 60.6 & - & - & 88.4 & 85.8 & 9.4 & - & - & 58.1 & - & - & 84 & 79.7 & 9.3 \\

\midrule
\multirow{5}{*}{16} 
& RD1 & 45 & 107 & 59.2 & 5 & 56 & 83.6 & 75.0 & 7.9 & 40 & 111 & 52.8 & 6 & 51 & 78.1 & 72.5 & 7.9\\ 
& RD2 &  52 & 124 & 59.6 & 4 & 51 & 85.9 & 78.4 & 11.1 & 47 & 127 & 54.1 & 3 & 46 & 76.6 & 73.5 & 10.4\\
& RD3 &  38 & 109 & 59.6 & 3 & 53 & 91.8 & 87.1 & 9.1 & 35 & 115 & 54.3 & 0 & 53 & 86.9 & 82.8 & 9.4\\
& RD4 & 76 & 157 & 65.9 & 4 & 81 & 92.4 & 90.7 & 9.6 & 74 & 141 & 66.2 & 7 & 76 & 90.2 & 89.3 & 8.9\\
\cmidrule{2-18}
&Average & - & - & 61.1 & - & - & 88.4 & 82.8 & 9.4 & - & - & 56.9 & - & - & 82.9 & 79.5 & 9.1 \\

\midrule
\multirow{5}{*}{42} 
& RD1 & 41 & 110 & 54.1 & 6 & 51 & 78.1 & 72.5 & 7.9 & 46 & 105 & 60.7 & 6 & 56 & 84.9 & 85.0 & 7.9\\ 
& RD2 & 52 & 131 & 58.7 & 3 & 52 & 85.9 & 82.3 & 10.9 & 46 & 124 & 53.5 & 3 & 46 & 76.6 & 71.5 & 10.2\\
& RD3 & 38 & 106 & 60.1 & 3 & 53 & 91.8 & 88.5 & 9.0 & 36 & 112 & 56.3 & 2 & 52 & 88.5 & 85.7 & 9.3\\
& RD4 & 75 & 156 & 65.2 & 4 & 80 & 65.2 & 90.7 & 9.6 & 75 & 138 & 67.3 & 10 & 74 & 91.3 & 90.7 & 8.9\\
\cmidrule{2-18}
&Average & - & - & 59.5 & - & - & 80.3 & 83.6 & 9.3 & - & - & 59.5 & - & - & 85.3 & 83.2 & 9.1 \\

\midrule
\multirow{5}{*}{256} 
& RD1 & 42 & 105 & 56.0 & 5 & 53 & 79.5 & 75.0 & 7.7 & 39 & 112 & 51.5 & 5 & 51 & 76.7 & 70.4 & 7.9\\ 
& RD2 & 52 & 135 & 58.2 & 3 & 51 & 84.4 & 81.8 & 11.2 & 49 & 137 & 54.9 & 1 & 50 & 79.7 & 74.0 & 11.1\\
& RD3 & 38 & 110 & 59.4 & 3 & 53 & 91.8 & 88.6 & 9.3 & 37 & 108 & 58.4 & 1 & 54 & 90.2 & 87.1 & 9.1\\
& RD4 & 76 & 156 & 66.0 & 4 & 81 & 92.4 & 91.3 & 9.6 & 74 & 162 & 63.8 & 7 & 76 & 90.2 & 89.3 & 9.8\\
\cmidrule{2-18}
&Average & - & - & 59.9 & - & - & 87.0 & 84.2 & 9.4 & - & - & 57.1 & - & - & 84.2 & 80.2 & 9.5 \\

\midrule
\multirow{5}{*}{\scriptsize{13374224}} 
& RD1 & 41 & 107 & 54.4 & 5 & 52 & 78.1 & 72.5 & 7.7 & 39 & 106 & 52.3 & 5 & 50 & 75.3 & 71.2 & 7.6\\ 
& RD2 & 52 & 129 & 59.0 & 3 & 52 & 85.9 & 83.8 & 10.8 & 52 & 132 & 58.6 & 3 & 52 & 85.9 & 82.3 & 11\\
& RD3 & 37 & 112 & 57.6 & 2 & 53 & 90.2 & 87.1 & 9.3 & 38 & 103 & 60.7 & 3 & 53 & 91.8 & 88.6 & 8.8\\
& RD4 & 76 & 153 & 66.3 & 4 & 81 & 92.4 & 92.1 & 9.5 & 73 & 157 & 63.6 & 5 & 77 & 89.1 & 88.7 & 9.6\\
\cmidrule{2-18}
&Average & - & - & 59.3 & - & - & 86.7 & 83.9 & 9.3 & - & - & 58.8 & - & - & 85.5 & 82.7 & 9.2 \\
\bottomrule
\end{tabularx}
 \end{threeparttable}
 \end{table*}


\sectopic{Evaluation Metrics.} Same as in RQ3. 

\sectopic{Results. } 
First, to investigate how \RICE performs under different random seed and temperature settings, we run the prompt with five distinct random seeds and two temperature values (0 and 0.5), the latter of which introduces additional randomness in the response. In other words, raising the temperature increases the model’s output entropy, meaning it becomes more likely to select lower-probability tokens. Table~\ref{tab:rq4-seed} shows the results of the \RICE approach, by prompting GPT-4o, using these settings. We observe that the average performance changes slightly when the temperature is held constant and the seed value is changed. The maximum changes are 2.6 pp in F2 score (the temperature is 0.5, and the seed changes from 16 to 42) and 8.1 pp in SR value (the temperature is 0, and the seed changes from 0 to 42).

In contrast, increasing randomness by setting the temperature equal to 0.5 caused a drop in F2 score for a constant seed value. The largest drop was observed for seed 16, from 61.1\% to 56.9\%. Moreover, we observe that the model’s behavior is not consistent across test documents. For example, when using a seed value of 13374224, increasing the temperature from 0 to 0.5 decreases the F2 score for RD1, RD2, and RD4, whereas it increases it for RD3. Nonetheless, we observe that \RICE shows little variation across different seeds when the temperature is set to 0, suggesting that random variation in results is primarily due to higher temperature. Thus, we select the performance at temperature 0 and arbitrarily set the seed to 16 to compare against the baselines.


\romina{
Moreover, we evaluated \RICE using different number of few-shot examples (3, 3, and 4) to assess the sensitivity of its performance to the degree of few-shot supervision.
We randomly chose 2, 3, and 4 examples from the existing few-shot examples. This created three different few-shot sets for each document. We ran the model separately with each set and recorded the resulting F2 scores.
Based on the results, the average F2 score across 2-, 3-, and 4-shot settings are 56.5\% (-4.6 pp), 58.9\% (-2.2 pp), and 59.5\% (-1.6 pp), respectively.
Nonetheless, \RICE consistently outperforms the baseline approaches across all example counts.
}



Table~\ref{tab:rq4-1} shows the results of the \RICE approach and the baseline prompts. \RICE outperforms all prompts at both the trace and requirement levels. 
\textit{P1} includes a retrieval phase that selects the top-k artifacts based on semantic similarity. In our case, the artifacts are the 26 GDPR provisions. Therefore, we experimented with two k values: 13 and 26, corresponding to half and all the GDPR provisions, respectively. Since using k=13 resulted in a 12.6\% loss of provisions during the retrieval step by the semantic similarity technique, we chose to include all provisions, given that the total number of provisions is not excessively large in our case.
\textit{P1} exhibits a lower number of retrieved links (TP+FP) compared to \RICE and \textit{P2}. Based on our observations of the model's rationale, this stems primarily from its inability to capture indirect relationships between requirements and their corresponding provisions. For instance, a requirement stating that a user's profile must include at least one charging point may seem purely functional but implicitly involves handling personal data, such as user identity or location. Consequently, it should be linked to the GDPR provision governing the obtaining of user consent for data processing. The prompt’s inability to capture this implicit connection results in lower retrieval performance and the omission of true trace links. Moreover, \textit{P1}’s F2 score is only 0.8~pp higher than \kashif (13.5\% in Table~\ref{tab:rq3}), which suggests that this prompt does not fully leverage GPT-4o’s capabilities, while \kashif already captures much of the underlying knowledge by being tuned over another dataset, allowing it to remain highly competitive.
\textit{P2} exhibits better performance, achieving a higher retrieval compared to \textit{P1}. We observed that the rationales generated by P2 provide more detail on how to link specific requirements to the relevant provisions. While \textit{P2}'s high retrieval rate increased the number of TPs, it concurrently increased the FPs. This ultimately led to a higher F2 score than \textit{P1}. \textit{P3\_1} is similar to \textit{P2} in terms of a low retrieval rate, while \textit{P3\_2} resulted in producing the highest FPs compared to all other prompt templates.  Moreover, \kashif has a higher recall compared to \textit{P3\_1}, and \textit{P3\_2} has a lower F2 score compared to \kashif.
On the other hand, the inclusion of context and detailed instructions in \RICE yields more effective outputs from GPT-4o, leading to improved performance at both the trace link and requirement levels. It is worth noting that \RICE, under all temperature and random seed configurations reported in Table~\ref{tab:rq4-seed}, consistently outperforms the baselines. Nonetheless, an investigation into which specific elements of the prompt design influence the model's behavior falls outside the scope of the current study.

\begin{table*}
\footnotesize
\centering
\caption{Accuracy of \RICE approach and baselines on the test datasets (RQ4).}
\label{tab:rq4-1}
\begin{threeparttable}[t]
\begin{tabularx}{\textwidth}{@{}p{0.11\textwidth} @{\hskip 0.5em} l*{20}{>{\centering\arraybackslash}X}@{}}
\toprule
&& & &  \multicolumn{5}{c}{\textit{Trace Link Level}} & \multicolumn{5}{c}{\textit{Requirement Level}} \\ 
\cmidrule(lr){5-9} \cmidrule(lr){10-14}
&& N & T$^*$ & TP & FP & R & P & F2 & EM & PM & SR & MR & C  \\
\midrule
\multirow{5}{*}{\RICE} 
& RD1 & 73 & 57 & 45 & 107 & 78.9 & 29.6 & 59.2 & 5 & 56 & 83.6 & 79.2 & 8.4 \\ 
&RD2 & 64  & 65 & 52 & 124 & 80.0 & 29.5 & 59.6 & 4 & 51 & 85.9 & 82.4 & 10.6 \\
&RD3 & 61  & 43 & 38 & 109 & 88.4 & 25.9 & 59.6 & 3 & 53 & 91.8 & 88.6 & 9.2 \\
&RD4 & 92 & 86 & 76 & 157 & 88.4 & 32.6 & 65.9 & 4 & 81 & 92.4 & 92.3 & 9.7 \\
\cmidrule{2-14}
&Average & - & - & - & - & 83.9 & 29.4 & 61.1 & - & - & 88.4 & 84.1 & 9.5 \\

\midrule
\multirow{5}{*}{\textit{P1}~\cite{hey2025requirements}} 
& RD1 & 73 & 57 & 16 & 25 & 28.1 & 39.0 & 29.7 & 31 & 5 & 49.3 & 18.7 & 2.1 \\ 
&RD2 & 64  & 65 & 15 & 46 & 23.3 & 24.6 & 23.4 & 31 & 6 & 57.8 & 23.0 & 3.7 \\
&RD3 & 61  & 43 & 1 & 8 & 2.3 & 11.1 & 2.8 & 25 & 0 & 40.9 & 1.4 & 0.5\\
&RD4 & 92 & 86 & 1 & 9 & 1.2 & 10.0 & 1.4 & 17 & 0 & 18.5 & 0.6 & 0.04\\
\cmidrule{2-14}
&Average & - & - & - & - & 13.7 & 21.2 & 14.3 & - & - & 41.6 & 13.4 & 1.6 \\

\midrule
\multirow{5}{*}{\textit{P2}~\cite{ronanki2024requirements}} 
& RD1 & 73 & 57 & 38 & 235 & 67.9 & 13.9 & 38.2 & 3 & 52 & 75.3 & 65.4 & 14.4\\ 
& RD2 & 64 & 65 & 45 & 475 & 69.2 & 8.6 & 28.8 & 0 & 49 & 76.6 & 71.0 & 31.2 \\ 
& RD3 & 61 & 43 & 19 & 291 & 44.2 & 6.1 & 19.7 & 0 & 37 & 60.7 & 19.5 & 19.5 \\ 
& RD4 & 92 & 86 & 36 & 525 & 41.9 & 6.4 & 19.9 & 1 & 42 & 46.7 & 37.0 & 23.4 \\ 
\cmidrule{2-14}
&Average & - & - & - & - & 55.8 & 8.7 & 26.6 & - & - & 64.8 & 48.2 & 22.1 \\

\midrule
\multirow{5}{*}{\textit{P3\_1}~\cite{ge2025cross}} 
& RD1 & 73 & 57 & 18 & 10 & 31.6 & 64.3 & 35.2 & 36 & 4 & 54.8 & 28.3 & 1.5\\ 
&RD2 & 64  & 65 & 9 & 30 & 13.8 & 32.1 & 15.1 & 29 & 3 & 50.0 & 13.2 & 2.4 \\
&RD3 & 61  & 43 & 5 & 11 & 11.6 & 31.2 & 13.3 & 26 & 1 & 44.3 & 8.5 & 1.0 \\
&RD4 & 92 & 86 & 1 & 4 & 1.2 & 20.0 & 1.4 & 19 & 1 & 21.7 & 0.6 & 0.2\\
\cmidrule{2-14}
&Average & - & - & - & - & 14.5 & 32.3 & 16.2 & - & - & 42.7 & 12.6 & 1.3 \\

\midrule
\multirow{5}{*}{\textit{P3\_2}~\cite{ge2025cross}} 
& RD1 & 73 & 57 & 26 & 445 & 45.6 & 5.5 & 18.6 & 1 & 47 & 65.7 & 43.3 & 24.8\\ 
&RD2 & 64  & 65 & 23 & 465 & 35.4 & 4.7 & 15.4 & 0 & 38 & 59.4 & 38.2 & 29.3\\
&RD3 & 61  & 43 & 7 & 413 & 16.3 & 1.7 & 5.9 & 1 & 29 & 49.1 & 14.3 & 26.4\\
&RD4 & 92 & 86 & 16 & 532 & 18.6 & 2.9 & 8.9 & 1 & 33 & 37.0 & 18.7 & 22.9\\
\cmidrule{2-14}
&Average & - & - & - & - & 29.0 & 3.7 & 12.2 & - & - & 52.8 & 28.6 & 25.8 \\

\bottomrule
\end{tabularx}
\begin{tablenotes}
     \it\item[] T$^*$: Predicted trace links, EM: Exact Match, PM: Partial Match, SR: Success Rate, MR: Macro Recall, C: Cost. 
     \end{tablenotes}
 \end{threeparttable}
 \end{table*}

\begin{figure}
\includegraphics[width=\textwidth]{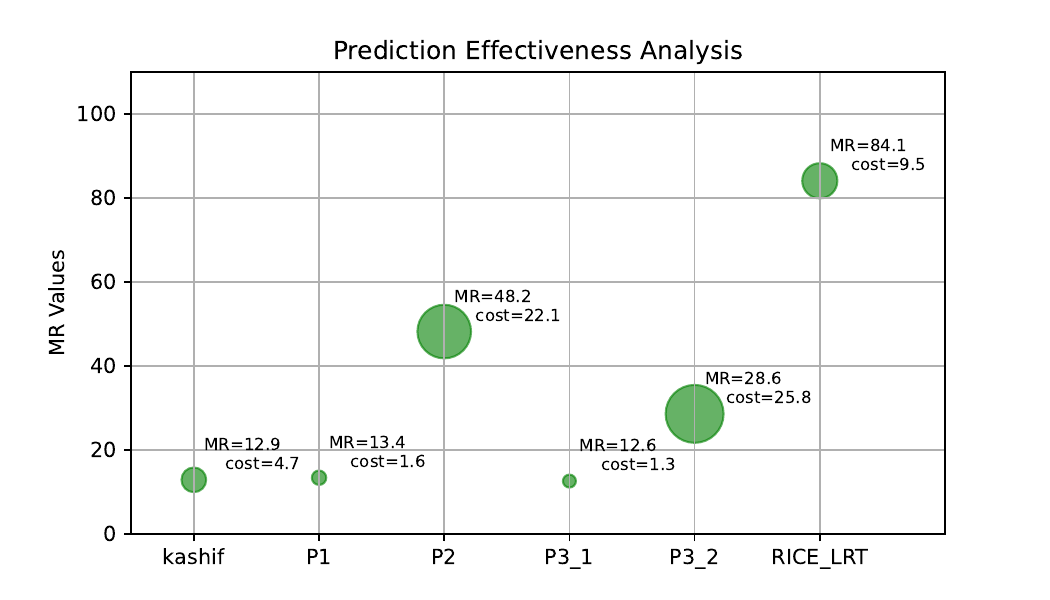}
  \centering
  \caption{\emph{MR} and \emph{Cost} comparison on the test documents (GDPR). The y-axis represents the average rate of successful trace-link predictions per requirement, while the bubble size indicates the total number of retrieved trace links. The larger the bubble, the more items retrieved from the entire set.}
  \label{fig:RQ4_mrcost}
\end{figure}

To more effectively illustrate how \RICE and the baselines perform on a per-requirement basis, we report their \emph{MR} and \emph{Cost} values. Figure~\ref{fig:RQ4_mrcost} depicts the performance of each prompt. The vertical position of each circle indicates the average recall for retrieving trace links for an individual requirement, while the circle’s radius reflects the total number of retrieved links, including both TPs and FPs. As visible, on average, \RICE achieves the highest recall at 84\% while retrieving only 9.5\% of all provisions. 


At the requirements level, the number of exact matches is very low for \RICE. We note that \RICE outputs at least one regulatory code for each requirement (based on our prompt of Section~\ref{subsec:Prompting_LLMs}) even when requirements do not have any trace links in the ground truth. This is one explanation for the sharp decrease in exact matches. Despite this, the number of partial matches has increased significantly, thereby improving the overall success rate. While an approach with a high exact-match rate would be ideal, the results remain beneficial, as discussed next. 

\begin{figure}
\includegraphics[width=0.9\textwidth]{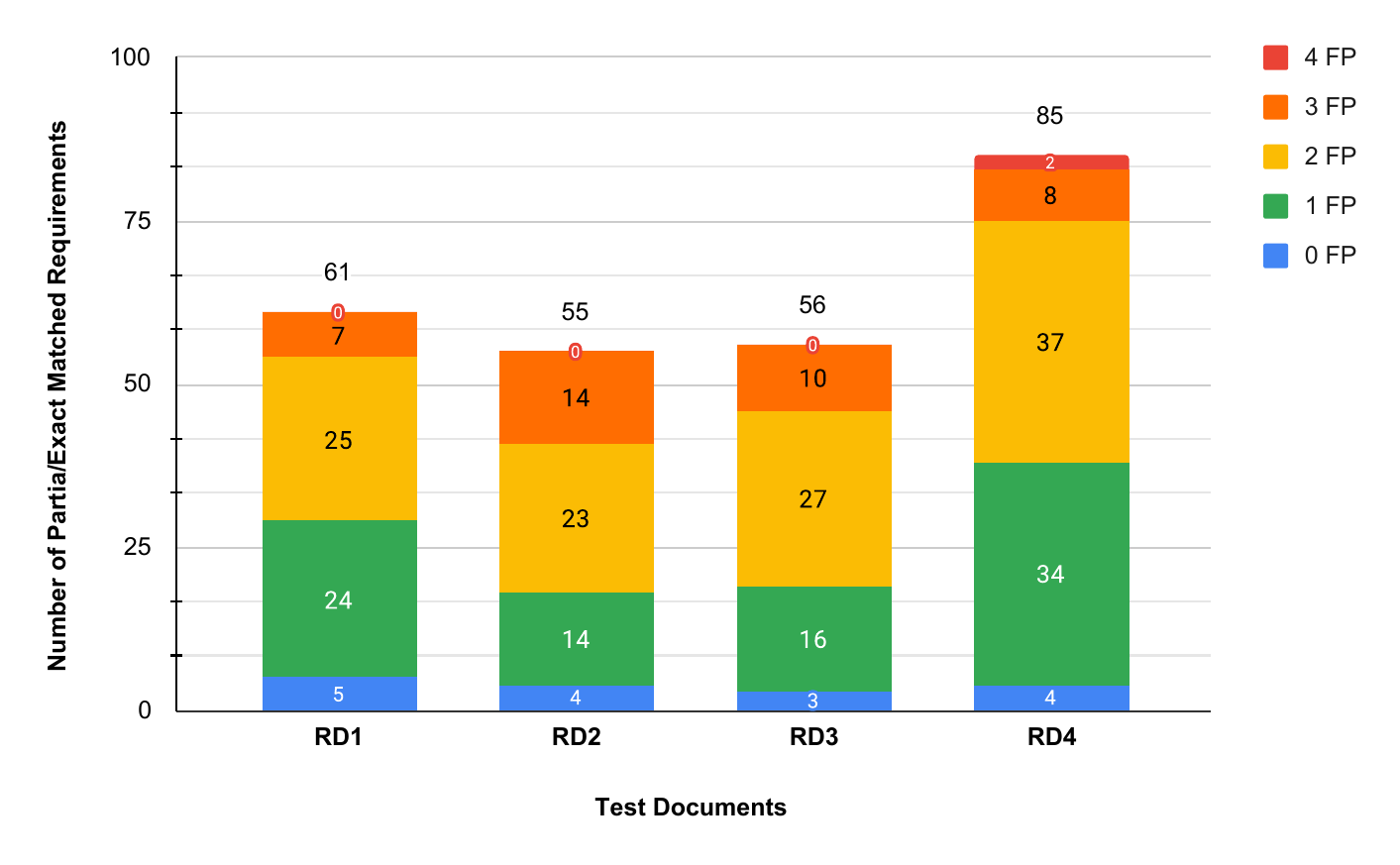}
  \centering
  \caption{Number of FPs for requirements with partial match produced by \RICE (Keepass: RD1, WASP: RD2, Datahub: RD3, and ScrumAlliance: RD4).}
  \label{fig:RQ4_FPsAnalysis}
\end{figure}

Fig.~\ref{fig:RQ4_FPsAnalysis} shows the split of partially matched requirements for the number of FPs. For instance, for RD1, 
there were 56 partially matched requirements. Of these, 24 (42.9\%) had only one FP, 25 (44.6\%) had two FPs, and the remaining 7 (12.5\%) had three FPs. As seen in the figure, across all four documents, there were very few requirements with four FPs. 
This indicates that most partially matched requirements had a manageable number of FPs, typically between 1 and 3. This result is significant because it suggests that the model's outputs are not overwhelming for analysts to process. Fewer FPs per requirement allow analysts to review and validate the proposed trace links more efficiently, reducing their cognitive load. Instead of starting from scratch or sifting through a vast set of 26 possible provisions per requirement, analysts can focus on validating and refining a much smaller, pre-filtered set of trace links. This aligns with the principle of assisted decision-making~\cite{skitka1999does}, in which automated tools augment human judgment by narrowing the set of options.


Our results further indicate that the GPT-4o model successfully demonstrated a stronger understanding of the LRT task when provided with an engineered prompt. This indicates that \RICE is effective at identifying the underlying logic and rationale behind provisions, even with a limited number of few-shot examples. Its ability to navigate complex relationships and extract logical links demonstrates its robustness in understanding the nuances of regulatory requirements. However, the cases it misses highlight areas in which the connections may require deeper, domain-specific knowledge or additional context to resolve ambiguities.

On investigating the FPs for each requirement, we observed that \rev{several predicted trace links may be relevant depending on the application context, even though they do not exactly match the ground truth. These false positives include provisions that are not in the ground truth but are relevant to the input requirements, as well as provisions in the ground truth (though not all of them), which is why they are considered partial matches. This underscores the potential of \RICE to identify trace links that indicate associations between requirements and provisions that may not have been considered during ground-truth construction. Such cases could still be informative to the analysts.}
For example, the \RICE output presented in Section~\ref{subsec:LLMQuerying} included three predictions with corresponding rationales. Of these, [SEC] is the ground truth, and [ACC] and [CNF] are categorized as FPs. The rationale for [ACC] highlights that requiring a key file ensures proper authentication, which can be interpreted as supporting the right to access. Similarly, the rationale for [CNF] emphasizes that protecting the database with a key file ensures sensitive data remains confidential. While these codes are not explicitly part of the ground truth for this requirement, they surface related regulatory considerations that may enrich the analyst's understanding of the requirement and its broader implications under GDPR. Hence, while FPs may not align perfectly with the ground truth, their contextual relevance, as inferred from the generated rationale, can offer valuable insights for the LRT task.
This also underscores the inherent subjectivity of the LRT task, especially when dealing with broadly framed regulations like GDPR, which often leave room for interpretation, compared to domain-specific regulations such as \texttt{HIPAA}. 

To assess the statistical significance of the differences in proportions of TPs versus FNs (recall) and TPs versus FPs (precision) between \kashif and \RICE, we employ Fisher’s exact test~\cite{fisher1922interpretation}. This test calculates the probability of obtaining the observed data (TPs, FPs, and FNs) under the null hypothesis
that proportions are equal across techniques. The p-values for RD1 for the proportions of (TPs, FPs) and (TPs, FNs) are 3e-4 and 1e-5, respectively. For RD2, the corresponding values are 2.5e-2 and 1e-5, for RD3, 4.6e-3 and 1e-5, and for RD4, 0 and 1e-5. The comparison between \RICE and \textit{P1}, , \textit{P3\_1}, and \textit{P3\_2} yields values comparable to those reported for \kashif, while the results for \textit{P2} are presented as follows: 0 and 5e-1, 1e-4 and 6e-1, 1e-4 and 5e-2, and 1e-4 and 3e-2 for RD1, RD2, RD3, and RD4, respectively. Across the four test documents, the resulting p-value is very small (below 5e-2). It falls well below the usual significance levels, indicating that the differences in precision and recall among \kashif, \textit{P1}, \textit{P2}, \textit{P3\_1}, \textit{P3\_2}, and \RICE are statistically significant in most cases, except for \textit{P2} and the recall scores for RD1 to RD2.

\begin{tcolorbox}[arc=1mm,width=\columnwidth,
                  top=0mm,left=0mm,  right=0mm, bottom=0mm,
                  boxrule=1pt, colback=violet!15!white,colframe=white, breakable]
\textbf{The answer to RQ4 is} that our \RICEORG-based approach, which utilizes an engineered prompting on GPT-4o, outperforms the baselines on the LRT task across the four test documents. 
Furthermore, the engineered context and instructions used in the few-shot setting of \RICE achieved the highest F2 score by maximizing TPs while minimizing FPs relative to the baselines. It also significantly outperforms \kashif, although \kashif remains competitive against \textit{P1} and \textit{P3\_2} prompts.
\end{tcolorbox}

\subsection{Discussion}\label{sec:discussion}
We experimented with a variety of classifiers, including two IR-, one ML-, two LM-, and one LLM-based methods. We compared them against our ST-based classifier, \kashif, on the LRT task using the publicly available HIPAA dataset. The results show that \kashif achieves the highest F2 score among all baselines, including \texttt{B}, TraceBERT, and LLaMA, which have been previously proposed for inter-requirements traceability. Moreover, \kashif achieved a strong overall ranking performance, obtaining an MAP of 81\% and the highest discrimination across thresholds, with an AUC of 0.93. \kashif demonstrated a superior ability to address the linguistic discrepancies between requirements and provisions, outperforming existing models in the literature that target traceability across different artifact types.
We also demonstrated that \RICE, as an engineered prompt, surpasses baseline prompts introduced in prior work, achieving a recall of 84\%, a precision of $\approx$30\%, and F2 score of 61\% on a newly created, more complex dataset using the GDPR provisions. From a more practical standpoint, we analyzed \RICE’s performance per requirement, using both Macro Recall (\emph{MR}) and \emph{Cost}. \RICE achieves an \emph{MR} of 84\% with a \emph{Cost} of 9.5\%, meaning that, on average, only 9.5\% of the total provisions are retrieved, while ensuring that 84\% of the true trace links are included in the retrieved list. This represents a 36-pp improvement in \emph{MR} over the best baseline prompt, while reducing \emph{Cost} by 13 pp. Moreover, we show that simple Yes/No–style prompts achieve low \emph{MR} and \emph{Cost} retrieval on the test documents, indicating that they do not fully leverage GPT-4o’s capabilities. Lastly, our \kashif classifier achieved competitive F2 scores on our newly created GDPR dataset compared to \textit{P1} and \textit{P3\_2} prompts, despite being fine-tuned only on the HIPAA dataset.

\section{{Threats to Validity}}~\label{sec:threats}
\sectopic{Internal Validity.} Bias is a well-known internal validity concern. To mitigate bias, for RQ3 and RQ4, the dataset comprising more than four documents was curated by two annotators with more than a decade of experience in RE.  Before the traceability sessions, there was no exposure to the technical details of our approach. The second potential threat to internal validity concerns the few-shot prompting in RQ4. The initial few-shot examples used for GPT-4o’s prompt engineering could introduce confirmation bias, potentially influencing the model’s predictions. To mitigate this, we designed the few-shot examples to reflect realistic usage scenarios in which an LLM serves as a recommendation tool guided by a human expert’s rationale for the first few requirements. This approach aligns with practical applications while minimizing the risk of confirmation bias. Additionally, the limited number of examples in the few-shot prompt was deliberately chosen to avoid overfitting. By doing so, we provided the LLM sufficient flexibility to independently apply reasoning to the remaining requirements, while maintaining a balance between guidance and adaptability. This approach ensures the LLM's outputs remain broadly applicable while minimizing potential validity threats, as seen by the relatively high success rate in RQ4.

\sectopic{External Validity.} We evaluated \kashif on two datasets, namely \texttt{HIPAA} and four new documents against GDPR. \texttt{HIPAA} is a pre-existing dataset frequently used in the RE literature. \rev{The test documents used in RQ3 and RQ4 (four new documents against GDPR),} which we created as part of our work, cover two types of textual requirements, including user stories and shall-type requirements. Such diversity helped increase the generalizability of our results. Experiments on more diverse requirements documents and other regulations are nonetheless required to improve the external validity of our study. \romina{Moreover, our evaluation is based on GPT-4o since this model is widely used in requirements engineering. Results obtained with other LLMs may differ; however, we expect them to be similar to those of LLMs with the same capabilities (focused on textual data and reasoning), and we leave a multi-model evaluation to future work.}

\section{Related Work}~\label{sec:related}
Requirements traceability (RT) has been extensively studied in RE~\cite{mucha2024systematic,li2023applications,tufail2017systematic,wang2018requirements,ramesh1998factors}. 
Existing work employs a range of technologies, from traditional methods such as Information Retrieval (IR) and statistical models to more advanced approaches such as Machine Learning (ML) and Deep Learning (DL). Early works borrowed IR techniques such as Vector Space Models (VSM), Latent Dirichlet Allocation (LDA), to find trace links between software artifacts via text relevancy~\cite{gao2022using, guo2017tackling, kuang2017analyzing, nishikawa2015recovering, panichella2013and, Mahmoud2013, Wang20193, sun2017frlink, capobianco2013improving, chhabra2017filtering, bavota2014enhancing, shao2013improved, Wang20191, Li2020,moran2020improving}.
More advanced techniques have been introduced using ML~\cite{Rasiman2022,Bella2019,mills2019tracing,Bella2018,Mills2018,Rath2018,Zhao2018,Mills2017,Mills20172,Falessi2017,Hayes2015,le2015rclinker,li2015recovering} and DL~\cite{kenton2019bert,Zhang2021,Wang20192,Guo:17,Chen2019,Alhoshan2019,Alhoshan2018,Li2018,zhao2020extended,Sultanov2013}, employing various algorithms — from classifiers like SVM, random forest, and decision trees to more sophisticated language models like BERT~\cite{devlin-etal-2019-bert} to find trace links. 
In recent years, with the emergence of LLMs, researchers have leveraged pre-trained knowledge via prompt engineering to identify trace links among software artifacts~\cite{hassine2024llm,rodriguez2023prompts,vogelsang2025impact}. Hassine~\cite{hassine2024llm} proposed an LLM-based technique that uses zero-shot learning on GPT3.5 to find trace links between requirements and goals in Goal-oriented Language (GRL) models. Moreover, Rodriguez et al.~\cite{rodriguez2023prompts} proposed an approach that integrates zero-shot prompting with reasoning to improve performance on the Traceability Link Recovery (TLR) problem across diverse software artifacts.
They have shown that a prompt that performs well with one model or dataset may not yield optimal results with another, underscoring the need to tailor prompts to the specific context. Recent studies have explored the use of prompting techniques for traceability within requirements~\cite{fuchss2025lissa,ronanki2024requirements,fuchss2025beyond,hey2025requirements}. Hey et al.~\cite{hey2025requirements} used Retrieval-Augmented Generation (RAG) with LLMs for inter-requirements traceability, where the model first retrieves relevant candidate requirements. Then, it identifies trace links among the retrieved candidates using the KISS prompt, which asks the model whether a trace link exists between a given pair of artifacts. Fuch{\ss}~\cite{fuchss2025beyond} et al. proposed an ensemble of LLM-based prompts for candidate filtering, where multiple LLM models are iteratively used to re-rank or filter the initial set of retrieved candidates. The prompt template used in this work follows the same design as that proposed by Hey et al.~\cite{hey2025requirements}. Additionally, Ronanki et al.~\cite{ronanki2024requirements} developed and evaluated five distinct interactive prompting patterns, each tuned to a specific dataset.

In addition to the algorithms being used, the types of artifacts with which these algorithms are intended to work also play a significant role. Existing studies primarily focus on identifying trace links between requirements and code~\cite{north2024code,peng2023enhancing,cleland:2010,panichella2013and,kuang2017analyzing,gao2022using,vogelsang2025impact}. Only a few studies have focused on establishing traceability across different software artifacts~\cite{nishikawa2015recovering}.
Existing approaches for RT are not directly applicable in our context due to the significant discrepancy between the legal language used in regulations and the technical language used in software requirements and related artifacts.  

Legal requirements traceability has only been investigated to a limited extent in the literature. Cleland-Huang et al.~\cite{cleland:2010} propose a probabilistic approach that identifies trace links between requirements and the \texttt{HIPAA} regulation by computing probabilities based on the detection of indicator terms for regulations that correspond to requirements. The authors further propose extending the indicator terms with more domain-specific terms retrieved from the web. In follow-up work, Gibiec et al.~\cite{Gibiec:2010} further investigate web mining. 
Guo et al.~\cite{Guo:17} extend the previous two papers to improve the terminology gap problem, i.e., the mismatch between terms in requirements and regulations. The authors investigate methods based on classification, ontologies, and web mining and evaluate them on \texttt{HIPAA}. 

While previous research has made significant strides in requirements traceability using traditional IR methods and ML/DL techniques, these approaches exhibit notable limitations in addressing the complexities of the LRT task. Most notably, existing methods struggle with the terminology gap between regulations and technical requirements, do not generalize well across regulations, and lack adaptability to multi-domain applications. Additionally, studies proposing prompt-based approaches have not fully leveraged the capabilities of prompt engineering, providing only detailed instructions and limited context. Their approaches primarily rely on pre- and post-processing strategies~\cite{hey2025requirements,fuchss2025beyond}, which, in some cases, query multiple large language models. 
In comparison to the above works, we empirically evaluate two automated LRT approaches: (1) a classifier-based solution leveraging sentence transformers and (2) a generative LLM-based solution guided by structured prompt engineering. By exploring these methods across two distinct regulations, \texttt{HIPAA} and GDPR, we advance the understanding of how modern NLP techniques can be adapted to meet the challenges of LRT. We also shed light on the possibilities, or lack thereof, of transfer learning across regulations. To the best of our knowledge, we are also among the first to identify the strengths and limitations of LLMs in this context. Further, larger studies involving human experts are required to establish the benefits of LLMs for LRT.

\section{Conclusion}~\label{sec:conclusion}
This study presents a comparative evaluation of two approaches to Legal Requirements Traceability (LRT): a classifier-based method, \kashif, leveraging Sentence Transformers, and a generative LLM-based method, \RICE, designed using a structured prompt engineering framework. 
Moreover, we compare \kashif against seven distinct baselines—ranging from traditional IR techniques to recent deep learning and transformer-based models (LSI, LDA, GloVe, TraceBERT, RoBERTa, and LLaMa2)—to comprehensively demonstrate its capabilities.
Our results demonstrate that \kashif provides significant improvements over the baselines in terms of F2 score, achieving a F2 score of $\approx$63\% on \texttt{HIPAA} data (23 pp more than the best baseline). 
On the other hand, \kashif's performance deteriorates on more complex datasets such as GDPR, yielding only 15\% recall. However, this also shows that \kashif, despite being tuned on a similar but less complex dataset, achieves performance comparable to a simple prompt (answering yes/no) on GPT-4o, which does not fully exploit its capabilities. 

Conversely, the \RICE approach, built on generative LLMs, outperformed \kashif on GDPR data, achieving an F2 score of 61\% and reducing manual effort for traceability by enabling analysts to vet only a fraction of trace links. Moreover, \RICE outperformed a simpler prompt, increasing TPs while reducing FPs.
These findings suggest that generative LLMs and carefully designed prompts offer a promising approach to automating LRT tasks in complex legal domains. However, the approach has its challenges, including the occurrence of false positives that require further investigation. In addition to evaluating current state-of-the-art methods, this work highlights critical challenges, including terminology gaps between requirements and provisions
By addressing these challenges, our study underscores the importance of tailoring solutions to the nuances of legal and regulatory contexts.

In the future, we plan to conduct a human-in-the-loop study with a domain expert to investigate the applicability of LLMs in the LRT context. We further plan to enhance LLM performance by incorporating domain-specific knowledge to better address terminology and contextual gaps between regulatory texts and technical requirements, particularly in the GDPR context.

\sectopic{Acknowledgement. }
This work is supported by the H2020 COSMOS European project, grant agreement No. 957254, NSERC of Canada under the Discovery and CRC programs, the Research Ireland grant 13/RC/2094-2, the Luxembourg National Research Fund under grant numbers C23\allowbreak/IS\allowbreak/17958091\allowbreak/PLAITO and NCER22/IS/16570468/NCER-FT. It is part of a collaborative research program between the University of Ottawa’s
Nanda laboratory and the SnT centre at the University of
Luxembourg. 


\section*{Declaration}
\textbf{Conflict of Interest.} The authors declared that they have no conflict of interest.

\bibliographystyle{IEEEtran}
\balance

\bibliography{paper}

\begin{thebibliography}{10}
\providecommand{\url}[1]{#1}
\csname url@samestyle\endcsname
\providecommand{\newblock}{\relax}
\providecommand{\bibinfo}[2]{#2}
\providecommand{\BIBentrySTDinterwordspacing}{\spaceskip=0pt\relax}
\providecommand{\BIBentryALTinterwordstretchfactor}{4}
\providecommand{\BIBentryALTinterwordspacing}{\spaceskip=\fontdimen2\font plus
\BIBentryALTinterwordstretchfactor\fontdimen3\font minus \fontdimen4\font\relax}
\providecommand{\BIBforeignlanguage}[2]{{%
\expandafter\ifx\csname l@#1\endcsname\relax
\typeout{** WARNING: IEEEtran.bst: No hyphenation pattern has been}%
\typeout{** loaded for the language `#1'. Using the pattern for}%
\typeout{** the default language instead.}%
\else
\language=\csname l@#1\endcsname
\fi
#2}}
\providecommand{\BIBdecl}{\relax}
\BIBdecl

\bibitem{Caruana:15}
R.~Caruana, Y.~Lou, J.~Gehrke, P.~Koch, M.~Sturm, and N.~Elhadad, ``Intelligible models for healthcare: Predicting pneumonia risk and hospital 30-day readmission,'' in \emph{21st ACM SIGKDD international conference on knowledge discovery and data mining}, 2015.

\bibitem{Zhan:22}
N.~Zhan, S.~Sarkadi, N.~C. Pacheco, and J.~Such, ``A model for governing information sharing in smart assistants,'' in \emph{AAAI/ACM Conference on AI, Ethics, and Society}.\hskip 1em plus 0.5em minus 0.4em\relax AAAI Press, 2022.

\bibitem{ahmad2023requirements}
K.~Ahmad, M.~Abdelrazek, C.~Arora, M.~Bano, and J.~Grundy, ``Requirements engineering for artificial intelligence systems: A systematic mapping study,'' \emph{Information and Software Technology}, p. 107176, 2023.

\bibitem{Feldt:18}
R.~Feldt, F.~G. de~Oliveira~Neto, and R.~Torkar, ``Ways of applying artificial intelligence in software engineering,'' in \emph{6th IEEE/ACM International Workshop on Realizing Artificial Intelligence Synergies in Software Engineering (RAISE)}, 2018.

\bibitem{GDPR}
\BIBentryALTinterwordspacing
{EU (GDPR)}, ``{Regulation (EU) 2016/679 of the European Parliament and of the Council of 27 April 2016 on the protection of natural persons with regard to the processing of personal data and on the free movement of such data, and repealing Directive 95/46/EC (General Data Protection Regulation), OJ L 119, 4.5.2016, p. 1–88},'' 2016. [Online]. Available: \url{http://data.europa.eu/eli/reg/2016/679/oj}
\BIBentrySTDinterwordspacing

\bibitem{Pohl:11}
P.~Klaus and R.~Chris, \emph{Requirements Engineering Fundamentals}, 1st~ed.\hskip 1em plus 0.5em minus 0.4em\relax Rocky Nook, 2011.

\bibitem{Meyer:22}
B.~Meyer, \emph{Handbook of Requirements and Business Analysis}.\hskip 1em plus 0.5em minus 0.4em\relax Springer Nature, 2022.

\bibitem{wang2018requirements}
B.~Wang, R.~Peng, Y.~Li, H.~Lai, and Z.~Wang, ``Requirements traceability technologies and technology transfer decision support: A systematic review,'' \emph{Journal of Systems and Software}, vol. 146, pp. 59--79, 2018.

\bibitem{tufail2017systematic}
H.~Tufail, M.~F. Masood, B.~Zeb, F.~Azam, and M.~W. Anwar, ``A systematic review of requirement traceability techniques and tools,'' in \emph{2017 2nd international conference on system reliability and safety (ICSRS)}.\hskip 1em plus 0.5em minus 0.4em\relax IEEE, 2017, pp. 450--454.

\bibitem{cleland:2010}
J.~Cleland-Huang, A.~Czauderna, M.~Gibiec, and J.~Emenecker, ``A machine learning approach for tracing regulatory codes to product specific requirements,'' in \emph{Proceedings of the 32nd ACM/IEEE International Conference on Software Engineering-Volume 1}, 2010, pp. 155--164.

\bibitem{Gibiec:2010}
M.~Gibiec, A.~Czauderna, and J.~Cleland-Huang, ``Towards mining replacement queries for hard-to-retrieve traces,'' in \emph{Proceedings of the 25th IEEE/ACM International Conference on Automated Software Engineering}, 2010, pp. 245--254.

\bibitem{Guo:17}
J.~Guo, M.~Gibiec, and J.~Cleland-Huang, ``Tackling the term-mismatch problem in automated trace retrieval,'' \emph{Empirical Software Engineering}, vol.~22, pp. 1103--1142, 2017.

\bibitem{deerwester1990indexing}
S.~Deerwester, S.~T. Dumais, G.~W. Furnas, T.~K. Landauer, and R.~Harshman, ``Indexing by latent semantic analysis,'' \emph{Journal of the American society for information science}, vol.~41, no.~6, pp. 391--407, 1990.

\bibitem{blei2003latent}
D.~M. Blei, A.~Y. Ng, and M.~I. Jordan, ``Latent dirichlet allocation,'' \emph{Journal of machine Learning research}, vol.~3, no. Jan, pp. 993--1022, 2003.

\bibitem{lin2021traceability}
J.~Lin, Y.~Liu, Q.~Zeng, M.~Jiang, and J.~Cleland-Huang, ``Traceability transformed: Generating more accurate links with pre-trained bert models,'' in \emph{2021 IEEE/ACM 43rd International Conference on Software Engineering (ICSE)}.\hskip 1em plus 0.5em minus 0.4em\relax IEEE, 2021, pp. 324--335.

\bibitem{ronanki2024requirements}
K.~Ronanki, B.~Cabrero-Daniel, J.~Horkoff, and C.~Berger, ``Requirements engineering using generative ai: Prompts and prompting patterns,'' in \emph{Generative AI for effective software development}.\hskip 1em plus 0.5em minus 0.4em\relax Springer, 2024, pp. 109--127.

\bibitem{hey2025requirements}
T.~Hey, D.~Fuch{\ss}, J.~Keim, and A.~Koziolek, ``Requirements traceability link recovery via retrieval-augmented generation,'' in \emph{International Working Conference on Requirements Engineering: Foundation for Software Quality}.\hskip 1em plus 0.5em minus 0.4em\relax Springer, 2025, pp. 381--397.

\bibitem{ge2025cross}
C.~Ge, T.~Wang, X.~Yang, and C.~Treude, ``Cross-level requirements tracing based on large language models,'' \emph{IEEE Transactions on Software Engineering}, 2025.

\bibitem{fuchss2025beyond}
D.~Fuch{\ss}, S.~Schwedt, J.~Keim, and T.~Hey, ``Beyond retrieval: A study of using llm ensembles for candidate filtering in requirements traceability,'' in \emph{2025 IEEE 33rd International Requirements Engineering Conference Workshops (RE)}, 2025.

\bibitem{Vaswani:17}
A.~Vaswani, N.~Shazeer, N.~Parmar, J.~Uszkoreit, L.~Jones, A.~N. Gomez, L.~Kaiser, and I.~Polosukhin, ``Attention is all you need,'' \emph{arXiv preprint arXiv:1706.03762}, 2017.

\bibitem{Reimers:19}
N.~Reimers and I.~Gurevych, ``Sentence-bert: Sentence embeddings using siamese bert-networks,'' \emph{CoRR}, vol. abs/1908.10084, 2019.

\bibitem{Jurafsky:20}
D.~Jurafsky and J.~H. Martin, \emph{Speech and Language Processing}, 3rd~ed.\hskip 1em plus 0.5em minus 0.4em\relax Pearson, 2020, \url{https://web.stanford.edu/~jurafsky/slp3/} (visited on 2022-01-04).

\bibitem{Alexandrescu:06}
A.~Alexandrescu and K.~Kirchhoff, ``Factored neural language models,'' in \emph{Proceedings of the Human Language Technology Conference of the North American Chapter of the Association for Computational Linguistics}, 2006, pp. 1--4.

\bibitem{Radford:18}
A.~Radford, K.~Narasimhan, T.~Salimans, I.~Sutskever \emph{et~al.}, ``Improving language understanding by generative pre-training,'' 2018.

\bibitem{touvron2023llama}
H.~Touvron, T.~Lavril, G.~Izacard, X.~Martinet, M.-A. Lachaux, T.~Lacroix, B.~Rozi{\`e}re, N.~Goyal, E.~Hambro, F.~Azhar \emph{et~al.}, ``Llama: Open and efficient foundation language models,'' \emph{arXiv preprint arXiv:2302.13971}, 2023.

\bibitem{touvron2023llama2}
H.~Touvron, L.~Martin, K.~Stone, P.~Albert, A.~Almahairi, Y.~Babaei, N.~Bashlykov, S.~Batra, P.~Bhargava, S.~Bhosale \emph{et~al.}, ``Llama 2: Open foundation and fine-tuned chat models,'' \emph{arXiv preprint arXiv:2307.09288}, 2023.

\bibitem{NEURIPS2020_1457c0d6}
T.~Brown, B.~Mann, N.~Ryder, M.~Subbiah, J.~D. Kaplan, P.~Dhariwal, A.~Neelakantan, P.~Shyam, G.~Sastry, A.~Askell, S.~Agarwal, A.~Herbert-Voss, G.~Krueger, T.~Henighan, R.~Child, A.~Ramesh, D.~Ziegler, J.~Wu, C.~Winter, C.~Hesse, M.~Chen, E.~Sigler, M.~Litwin, S.~Gray, B.~Chess, J.~Clark, C.~Berner, S.~McCandlish, A.~Radford, I.~Sutskever, and D.~Amodei, ``Language models are few-shot learners,'' in \emph{Advances in Neural Information Processing Systems}, H.~Larochelle, M.~Ranzato, R.~Hadsell, M.~Balcan, and H.~Lin, Eds., vol.~33.\hskip 1em plus 0.5em minus 0.4em\relax Curran Associates, Inc., 2020, pp. 1877--1901.

\bibitem{Goodfellow:16}
I.~Goodfellow, Y.~Bengio, and A.~Courville, \emph{Deep Learning}, 1st~ed.\hskip 1em plus 0.5em minus 0.4em\relax MIT Press, 2016.

\bibitem{Manning:08}
C.~Manning, P.~Raghavan, and H.~Schutze, \emph{Introduction to Information Retrieval}, 1st~ed.\hskip 1em plus 0.5em minus 0.4em\relax Cambridge University Press, 2008.

\bibitem{Yao2014}
X.~Yao, J.~Berant, and B.~Van~Durme, ``Freebase qa: Information extraction or semantic parsing?'' in \emph{Proceedings of the ACL 2014 Workshop on Semantic Parsing}, 2014, pp. 82--86.

\bibitem{Corley2005}
C.~D. Corley and R.~Mihalcea, ``Measuring the semantic similarity of texts,'' in \emph{Proceedings of the ACL workshop on empirical modeling of semantic equivalence and entailment}, 2005, pp. 13--18.

\bibitem{Amaral:21}
O.~Amaral, S.~Abualhaija, D.~Torre, M.~Sabetzadeh, and L.~Briand, ``{AI}-enabled automation for completeness checking of privacy policies,'' \emph{IEEE Transactions on Software Engineering}, vol.~48, no.~11, 2021.

\bibitem{vogelsang2024using}
A.~Vogelsang and J.~Fischbach, ``Using large language models for natural language processing tasks in requirements engineering: A systematic guideline,'' \emph{arXiv preprint arXiv:2402.13823}, 2024.

\bibitem{vogelsang2024specifications}
A.~Vogelsang, ``From specifications to prompts: On the future of generative large language models in requirements engineering,'' \emph{IEEE Software}, vol.~41, no.~5, pp. 9--13, 2024.

\bibitem{huang2025prompt}
K.~Huang, F.~Wang, Y.~Huang, and C.~Arora, ``Prompt engineering for requirements engineering: A literature review and roadmap,'' \emph{arXiv preprint arXiv:2507.07682}, 2025.

\bibitem{Amaral:23a}
O.~Amaral, M.~I. Azeem, S.~Abualhaija, and L.~Briand, ``{NLP}-based automated compliance checking of data processing agreements against {GDPR},'' \emph{IEEE Transactions on Software Engineering}, vol.~49, 2023.

\bibitem{annex}
R.~Etezadi, S.~Abualhaija, C.~Arora, and L.~Briand, \emph{``Online Annex (online)''}, 2024, available at \url{https://figshare.com/s/769b925bd426285e463b}, January 2024.

\bibitem{pennington2014glove}
J.~Pennington, R.~Socher, and C.~D. Manning, ``Glove: Global vectors for word representation,'' in \emph{Proceedings of the 2014 conference on empirical methods in natural language processing (EMNLP)}, 2014, pp. 1532--1543.

\bibitem{liu2019roberta}
Y.~Liu, M.~Ott, N.~Goyal, J.~Du, M.~Joshi, D.~Chen, O.~Levy, M.~Lewis, L.~Zettlemoyer, and V.~Stoyanov, ``Roberta: A robustly optimized bert pretraining approach,'' \emph{arXiv preprint arXiv:1907.11692}, 2019.

\bibitem{guo2025natural}
J.~L. Guo, J.-P. Stegh{\"o}fer, A.~Vogelsang, and J.~Cleland-Huang, ``Natural language processing for requirements traceability,'' in \emph{Handbook on Natural Language Processing for Requirements Engineering}.\hskip 1em plus 0.5em minus 0.4em\relax Springer, 2025, pp. 89--116.

\bibitem{baeza1999modern}
R.~Baeza-Yates, B.~Ribeiro-Neto \emph{et~al.}, \emph{Modern information retrieval}.\hskip 1em plus 0.5em minus 0.4em\relax ACM press New York, 1999, vol. 463, no. 1999.

\bibitem{husain2019codesearchnet}
H.~Husain, H.-H. Wu, T.~Gazit, M.~Allamanis, and M.~Brockschmidt, ``{CodeSearchNet} challenge: Evaluating the state of semantic code search,'' \emph{arXiv preprint arXiv:1909.09436}, 2019.

\bibitem{uto2017diverse}
M.~Uto, S.~Louvign{\'e}, Y.~Kato, T.~Ishii, and Y.~Miyazawa, ``Diverse reports recommendation system based on latent dirichlet allocation,'' \emph{Behaviormetrika}, vol.~44, no.~2, pp. 425--444, 2017.

\bibitem{silva2021topic}
C.~C. Silva, M.~Galster, and F.~Gilson, ``Topic modeling in software engineering research,'' \emph{Empirical Software Engineering}, vol.~26, no.~6, p. 120, 2021.

\bibitem{AroraHH24}
C.~Arora, T.~Herda, and V.~Homm, ``Generating test scenarios from {NL} requirements using retrieval-augmented llms: An industrial study,'' in \emph{32nd {IEEE} International Requirements Engineering Conference, {RE} 2024}.\hskip 1em plus 0.5em minus 0.4em\relax {IEEE}, 2024, pp. 240--251.

\bibitem{abs_2310_18648}
\BIBentryALTinterwordspacing
A.~Nguyen{-}Duc, B.~C. Daniel, A.~Przybylek, C.~Arora, D.~Khanna, T.~Herda, U.~Rafiq, J.~Melegati, E.~Guerra, K.~Kemell, M.~Saari, Z.~Zhang, H.~Le, T.~Quan, and P.~Abrahamsson, ``Generative artificial intelligence for software engineering - {A} research agenda,'' \emph{CoRR}, vol. abs/2310.18648, 2023. [Online]. Available: \url{https://doi.org/10.48550/arXiv.2310.18648}
\BIBentrySTDinterwordspacing

\bibitem{skitka1999does}
L.~J. Skitka, K.~L. Mosier, and M.~Burdick, ``Does automation bias decision-making?'' \emph{International Journal of Human-Computer Studies}, vol.~51, no.~5, pp. 991--1006, 1999.

\bibitem{fisher1922interpretation}
R.~A. Fisher, ``On the interpretation of $\chi$ 2 from contingency tables, and the calculation of p,'' \emph{Journal of the royal statistical society}, vol.~85, no.~1, pp. 87--94, 1922.

\bibitem{mucha2024systematic}
J.~Mucha, A.~Kaufmann, and D.~Riehle, ``A systematic literature review of pre-requirements specification traceability,'' \emph{Requirements Engineering}, pp. 1--23, 2024.

\bibitem{li2023applications}
X.~Li, B.~Wang, H.~Wan, Y.~Deng, and Z.~Wang, ``Applications of machine learning in requirements traceability: A systematic mapping study (s).'' in \emph{SEKE}, 2023, pp. 566--571.

\bibitem{ramesh1998factors}
B.~Ramesh, ``Factors influencing requirements traceability practice,'' \emph{Communications of the ACM}, vol.~41, no.~12, pp. 37--44, 1998.

\bibitem{gao2022using}
H.~Gao, H.~Kuang, K.~Sun, X.~Ma, A.~Egyed, P.~M{\"a}der, G.~Rong, D.~Shao, and H.~Zhang, ``Using consensual biterms from text structures of requirements and code to improve ir-based traceability recovery,'' in \emph{37th IEEE/ACM International Conference on Automated Software Engineering}, 2022, pp. 1--1.

\bibitem{guo2017tackling}
J.~Guo, M.~Gibiec, and J.~Cleland-Huang, ``Tackling the term-mismatch problem in automated trace retrieval,'' \emph{Empirical Software Engineering}, vol.~22, pp. 1103--1142, 2017.

\bibitem{kuang2017analyzing}
H.~Kuang, J.~Nie, H.~Hu, P.~Rempel, J.~L{\"u}, A.~Egyed, and P.~M{\"a}der, ``Analyzing closeness of code dependencies for improving ir-based traceability recovery,'' in \emph{2017 IEEE 24th International Conference on Software Analysis, Evolution and Reengineering (SANER)}.\hskip 1em plus 0.5em minus 0.4em\relax IEEE, 2017, pp. 68--78.

\bibitem{nishikawa2015recovering}
K.~Nishikawa, H.~Washizaki, Y.~Fukazawa, K.~Oshima, and R.~Mibe, ``Recovering transitive traceability links among software artifacts,'' in \emph{2015 IEEE International Conference on Software Maintenance and Evolution (ICSME)}.\hskip 1em plus 0.5em minus 0.4em\relax IEEE, 2015, pp. 576--580.

\bibitem{panichella2013and}
A.~Panichella, C.~McMillan, E.~Moritz, D.~Palmieri, R.~Oliveto, D.~Poshyvanyk, and A.~De~Lucia, ``When and how using structural information to improve ir-based traceability recovery,'' in \emph{2013 17th European Conference on Software Maintenance and Reengineering}.\hskip 1em plus 0.5em minus 0.4em\relax IEEE, 2013, pp. 199--208.

\bibitem{Mahmoud2013}
A.~Mahmoud and N.~Niu, ``Supporting requirements traceability through refactoring,'' in \emph{21st IEEE International Requirements Engineering Conference}, 2013, pp. 32--41.

\bibitem{Wang20193}
B.~Wang, R.~Peng, Z.~Wang, and Y.~Zhao, ``Combining vsm and btm to improve requirements trace links generation,'' in \emph{Proceedings of the International Conference on Software Engineering and Knowledge Engineering}, vol. 2019-July, 2019, pp. 567--572.

\bibitem{sun2017frlink}
Y.~Sun, Q.~Wang, and Y.~Yang, ``Frlink: Improving the recovery of missing issue-commit links by revisiting file relevance,'' \emph{Information and Software Technology}, vol.~84, pp. 33--47, 2017.

\bibitem{capobianco2013improving}
G.~Capobianco, A.~D. Lucia, R.~Oliveto, A.~Panichella, and S.~Panichella, ``Improving ir-based traceability recovery via noun-based indexing of software artifacts,'' \emph{Journal of Software: Evolution and Process}, vol.~25, no.~7, pp. 743--762, 2013.

\bibitem{chhabra2017filtering}
Jyoti and J.~K. Chhabra, ``Filtering of false positives from ir-based traceability links among software artifacts,'' in \emph{2nd International Conference for Convergence in Technology}, vol. 2017-January, 2017, pp. 1111--1115.

\bibitem{bavota2014enhancing}
G.~Bavota, A.~De~Lucia, R.~Oliveto, and G.~Tortora, ``Enhancing software artefact traceability recovery processes with link count information,'' \emph{Information and Software Technology}, vol.~56, no.~2, pp. 163--182, 2014.

\bibitem{shao2013improved}
J.~Shao, W.~Wu, and P.~Geng, ``An improved approach to the recovery of traceability links between requirement documents and source codes based on latent semantic indexing,'' in \emph{Computational Science and Its Applications--ICCSA 2013: 13th International Conference, Ho Chi Minh City, Vietnam, June 24-27, 2013, Proceedings, Part V 13}.\hskip 1em plus 0.5em minus 0.4em\relax Springer, 2013, pp. 547--557.

\bibitem{Wang20191}
S.~Wang, T.~Li, and Z.~Yang, ``Exploring semantics of software artifacts to improve requirements traceability recovery: A hybrid approach,'' in \emph{Asia-Pacific Software Engineering Conference}, vol. 2019-December, 2019, pp. 39--46.

\bibitem{Li2020}
T.~Li, S.~Wang, D.~Lillis, and Z.~Yang, ``Combining machine learning and logical reasoning to improve requirements traceability recovery,'' \emph{Applied Sciences (Switzerland)}, vol.~10, pp. 1--23, 2020.

\bibitem{moran2020improving}
K.~Moran, D.~N. Palacio, C.~Bernal-C{\'a}rdenas, D.~McCrystal, D.~Poshyvanyk, C.~Shenefiel, and J.~Johnson, ``Improving the effectiveness of traceability link recovery using hierarchical bayesian networks,'' in \emph{Proceedings of the ACM/IEEE 42nd International Conference on Software Engineering}, 2020, pp. 873--885.

\bibitem{Rasiman2022}
R.~Rasiman, F.~Dalpiaz, and S.~Espana, ``How effective is automated trace link recovery in model-driven development?'' in \emph{Requirements Engineering: Foundation for Software Quality, REFSQ 2022}, vol. 13216, 2022, pp. 35--51.

\bibitem{Bella2019}
E.~E. Bella, S.~Creff, M.-P. Gervais, and R.~Bendraou, ``Atlas: A framework for traceability links recovery combining information retrieval and semi-supervised techniques,'' in \emph{23rd International Enterprise Distributed Object Computing Conference}, 2019, pp. 161--170.

\bibitem{mills2019tracing}
C.~Mills, J.~Escobar-Avila, A.~Bhattacharya, G.~Kondyukov, S.~Chakraborty, and S.~Haiduc, ``Tracing with less data: active learning for classification-based traceability link recovery,'' in \emph{2019 IEEE International Conference on Software Maintenance and Evolution (ICSME)}.\hskip 1em plus 0.5em minus 0.4em\relax IEEE, 2019, pp. 103--113.

\bibitem{Bella2018}
E.~E. Bella, M.-P. Gervais, R.~Bendraou, L.~Wouters, and A.~Koudri, ``Semi-supervised approach for recovering traceability links in complex systems,'' in \emph{IEEE International Conference on Engineering of Complex Computer Systems}, vol. 2018-December, 2018, pp. 193--196.

\bibitem{Mills2018}
C.~Mills, J.~Escobar-Avila, and S.~Haiduc, ``Automatic traceability maintenance via machine learning classification,'' in \emph{IEEE International Conference on Software Maintenance and Evolution, ICSME 2018}.\hskip 1em plus 0.5em minus 0.4em\relax Institute of Electrical and Electronics Engineers Inc., 11 2018, pp. 369--380.

\bibitem{Rath2018}
M.~Rath, J.~Rendall, J.~L. Guo, J.~Cleland-Huang, and P.~Mäder, ``Traceability in the wild: Automatically augmenting incomplete trace links,'' in \emph{International Conference on Software Engineering}.\hskip 1em plus 0.5em minus 0.4em\relax IEEE Computer Society, 5 2018, pp. 834--845.

\bibitem{Zhao2018}
T.~Zhao, Q.~Cao, and Q.~Sun, ``An improved approach to traceability recovery based on word embeddings,'' in \emph{Proceedings - Asia-Pacific Software Engineering Conference, APSEC}, vol. 2017-December, 2018, pp. 81--89.

\bibitem{Mills2017}
C.~Mills, ``Towards the automatic classification of traceability links,'' in \emph{IEEE/ACM International Conference on Automated Software Engineering}, 2017, pp. 1018--1021.

\bibitem{Mills20172}
C.~Mills and S.~Haiduc, ``A machine learning approach for determining the validity of traceability links,'' in \emph{IEEE/ACM International Conference on Software Engineering Companion, ICSE-C 2017}, 2017, pp. 121--123.

\bibitem{Falessi2017}
D.~Falessi, M.~D. Penta, G.~Canfora, and G.~Cantone, ``Estimating the number of remaining links in traceability recovery,'' \emph{Empirical Software Engineering}, vol.~22, pp. 996--1027, 6 2017.

\bibitem{Hayes2015}
J.~H. Hayes, G.~Antoniol, B.~Adams, and Y.-G. Guehénéuc, ``Inherent characteristics of traceability artifacts: Less is more,'' in \emph{23rd IEEE International Requirements Engineering Conference, RE 2015 - Proceedings}, 2015, pp. 196--201.

\bibitem{le2015rclinker}
T.-D.~B. Le, M.~Linares-V{\'a}squez, D.~Lo, and D.~Poshyvanyk, ``Rclinker: Automated linking of issue reports and commits leveraging rich contextual information,'' in \emph{2015 IEEE 23rd international conference on program comprehension}.\hskip 1em plus 0.5em minus 0.4em\relax IEEE, 2015, pp. 36--47.

\bibitem{li2015recovering}
Z.~Li, M.~Chen, L.~Huang, and V.~Ng, ``Recovering traceability links in requirements documents,'' in \emph{Proceedings of the Nineteenth Conference on Computational Natural Language Learning}, 2015, pp. 237--246.

\bibitem{kenton2019bert}
J.~D. M.-W.~C. Kenton and L.~K. Toutanova, ``Bert: Pre-training of deep bidirectional transformers for language understanding,'' in \emph{Proceedings of NAACL-HLT}, 2019, pp. 4171--4186.

\bibitem{Zhang2021}
M.~Zhang, C.~Tao, H.~Guo, and Z.~Huang, ``Recovering semantic traceability between requirements and source code using feature representation techniques,'' in \emph{IEEE International Conference on Software Quality, Reliability and Security, QRS}, vol. 2021-December, 2021, pp. 873--882.

\bibitem{Wang20192}
S.~Wang, T.~Li, and Z.~Yang, ``Using graph embedding to improve requirements traceability recovery,'' in \emph{Applied Informatics: Second International Conference, ICAI 2019, Madrid, Spain, November 7--9, 2019, Proceedings 2}.\hskip 1em plus 0.5em minus 0.4em\relax Springer, 2019, pp. 533--545.

\bibitem{Chen2019}
L.~Chen, D.~Wang, J.~Wang, and Q.~Wang, ``Enhancing unsupervised requirements traceability with sequential semantics,'' in \emph{Proceedings - Asia-Pacific Software Engineering Conference, APSEC}, vol. 2019-December, 2019, pp. 23--30.

\bibitem{Alhoshan2019}
W.~Alhoshan, R.~Batista-Navarro, and L.~Zhao, ``Using frame embeddings to identify semantically related software requirements,'' in \emph{CEUR Workshop Proceedings}, vol. 2376, 2019.

\bibitem{Alhoshan2018}
W.~Alhoshan, L.~Zhao, and R.~Batista-Navarro, ``Using semantic frames to identify related textual requirements: An initial validation,'' in \emph{International Symposium on Empirical Software Engineering and Measurement}, 2018.

\bibitem{Li2018}
Y.~Li, S.~Schulze, and G.~Saake, ``Extracting features from requirements: Achieving accuracy and automation with neural networks,'' in \emph{25th IEEE International Conference on Software Analysis, Evolution and Reengineering, SANER 2018 - Proceedings}, vol. 2018-March, 2018, pp. 477--481.

\bibitem{zhao2020extended}
G.~Zhao, T.~Li, and Z.~Yang, ``An extended knowledge representation learning approach for context-based traceability link recovery,'' in \emph{2020 IEEE Seventh International Workshop on Artificial Intelligence for Requirements Engineering (AIRE)}.\hskip 1em plus 0.5em minus 0.4em\relax IEEE, 2020, pp. 22--22.

\bibitem{Sultanov2013}
H.~Sultanov and J.~H. Hayes, ``Application of reinforcement learning to requirements engineering: Requirements tracing,'' in \emph{21st IEEE International Requirements Engineering Conference}, 2013, pp. 52--61.

\bibitem{devlin-etal-2019-bert}
J.~Devlin, M.-W. Chang, K.~Lee, and K.~Toutanova, ``{BERT}: Pre-training of deep bidirectional transformers for language understanding,'' in \emph{Proceedings of the 2019 Conference of the North {A}merican Chapter of the Association for Computational Linguistics: Human Language Technologies, Volume 1 (Long and Short Papers)}, J.~Burstein, C.~Doran, and T.~Solorio, Eds.\hskip 1em plus 0.5em minus 0.4em\relax Minneapolis, Minnesota: Association for Computational Linguistics, Jun. 2019, pp. 4171--4186.

\bibitem{hassine2024llm}
J.~Hassine, ``An llm-based approach to recover traceability links between security requirements and goal models,'' in \emph{Proceedings of the 28th International Conference on Evaluation and Assessment in Software Engineering}, 2024, pp. 643--651.

\bibitem{rodriguez2023prompts}
A.~D. Rodriguez, K.~R. Dearstyne, and J.~Cleland-Huang, ``Prompts matter: Insights and strategies for prompt engineering in automated software traceability,'' in \emph{2023 IEEE 31st International Requirements Engineering Conference Workshops (REW)}.\hskip 1em plus 0.5em minus 0.4em\relax IEEE, 2023, pp. 455--464.

\bibitem{vogelsang2025impact}
A.~Vogelsang, A.~Korn, G.~Broccia, A.~Ferrari, J.~Fischbach, and C.~Arora, ``On the impact of requirements smells in prompts: The case of automated traceability,'' in \emph{IEEE/ACM 50th international conference on software engineering: new ideas and emerging results (ICSE-NIER)}, 2025.

\bibitem{fuchss2025lissa}
D.~Fuch{\ss}, T.~Hey, J.~Keim, H.~Liu, N.~Ewald, T.~Thirolf, and A.~Koziolek, ``Lissa: toward generic traceability link recovery through retrieval-augmented generation,'' in \emph{Proceedings of the IEEE/ACM 47th International Conference on Software Engineering. ICSE}, vol.~25, 2025.

\bibitem{north2024code}
M.~North, A.~Atapour-Abarghouei, and N.~Bencomo, ``Code gradients: Towards automated traceability of llm-generated code,'' in \emph{2024 IEEE 32nd International Requirements Engineering Conference (RE)}.\hskip 1em plus 0.5em minus 0.4em\relax IEEE, 2024, pp. 321--329.

\bibitem{peng2023enhancing}
T.~Peng, K.~She, Y.~Shen, X.~Xu, and Y.~Yu, ``Enhancing traceability link recovery with fine-grained query expansion analysis,'' \emph{Information}, vol.~14, no.~5, p. 270, 2023.

\end{thebibliography}

\end{document}